\documentclass[12pt]{iopart}

\usepackage{graphicx}
\usepackage{amsmath}
\usepackage{amssymb}
\usepackage{hyperref}
\usepackage{caption}
\usepackage{xcolor}

\begin{document}

\title{Logarithmic catastrophes and Stokes's phenomenon in waves at horizons}

\author{LM Farrell$^1$, CJ Howls$^2$, DHJ O'Dell$^1$}

\address{$^1$ Department of Physics and Astronomy, McMaster University, 1280 Main St. W., Hamilton, Ontario, Canada L8S 4M1}
\address{$^2$ Mathematical Sciences, University of Southampton, Southampton, SO17 1BJ, UK}

\ead{dodell@mcmaster.ca}
\ead{c.j.howls@soton.ac.uk}
\vspace{10pt}


\begin{abstract}
Waves propagating near an event horizon display interesting features including logarithmic phase singularities and caustics. We consider an acoustic horizon in a flowing Bose-Einstein condensate where the elementary excitations obey the Bogoliubov dispersion relation. In the hamiltonian ray theory the solutions undergo a broken pitchfork bifurcation near the horizon and one might therefore expect the associated wave structure to be given by a Pearcey function, this being the universal wave function that dresses catastrophes with two control parameters. However, the wave function is in fact an Airy-type function supplemented by a logarithmic phase term, a novel type of wave catastrophe. Similar wave functions arise in aeroacoustic flows from jet engines, path integrals in radio astronomy, and also gravitational horizons if dispersion which violates Lorentz symmetry in the UV is included. The approach we take differs from most previous authors in that we analyze the behaviour of the integral representation of the wave function using exponential coordinates. This allows for a different treatment of the branch cuts and gives rise to an analysis based purely on saddlepoint expansions. We are thereby able to resolve the multiple real and complex waves that interact at the horizon and its companion caustic. We find that the horizon is a physical manifestation of a Stokes surface, marking the place where a wave is born, and that the horizon and the caustic do not in general coincide: the finite spatial region between them delineates a broadened horizon. 
\end{abstract}

%
%
%
%
%

\section{Dedication}

This paper is dedicated to Sir Michael Berry in celebration of his 80th birthday. Of his many contributions to physics and physical asymptotics, one of the major themes of his work is his interest in waves near singularities, ranging from the most dramatic occurrences such as tsunamis \cite{berry376_2005,berry399_2007} and tidal bores \cite{berry505_2018,berry513_2019}, to the most gentle (yet profound) in the form of Stokes's phenomenon \cite{berry181_1989,berry190_1989,berry197_1990,BerryHowls}.  In particular, he has devised minimal models for undular bores that reveal in a characteristically clear way the central role played by caustics as well as an analogy to the Hawking effect \cite{berry505_2018}, and it is a related connection we pick up here in the context of a flowing superfluid. Starting in the 1970s \cite{berry49_1976} and continuing today \cite{berry531_2021}, Michael Berry has championed the application of catastrophe theory to caustics and we humbly follow in his footsteps in this paper, focusing on novel catastrophes with logarithmic phase singularities that accompany an acoustic event horizon in a superfluid.

\section{Introduction}

If the flow speed of a fluid exceeds the speed of waves in the fluid then the latter are unable to propagate against the flow and an analogue of an event horizon occurs. This situation is capable of mimicking many aspects of black hole physics where the effective spacetime metric depends on the fluid flow  and acoustic waves play the role of light \cite{unruh81,unruh95,schutzhold2002,leonhardt2003,barcelo2011}. Analogue event horizons for classical waves, including the classical analogue of Hawking radiation, have been   observed in water tank experiments \cite{Rousseaux2008,Rousseaux2010,weinfurter2011,Euve2016,euve2020,euve2021,fourdrinoy2022} and also in optical fibres \cite{philbin2008,Belgiorno2010,Drori2019}. Another system where analogue event horizons can be created is a flowing Bose-Einstein condensate (BEC) formed from ultracold atoms. These are among the simplest examples of superfluids and are so cold that quantum processes such as the analogue of spontaneous Hawking radiation can dominate thermally activated phonons. Analogue Hawking radiation in BECs has been anticipated theoretically for over twenty years \cite{leonhardt2003,garay2000,garay2001,leonhardt03b,balbinot08,carusotto08,macher09,recati09,mayoral2011,larre12,nova14}, and was recently realized in a series of experiments by J. Steinhauer and coworkers \cite{lahav10,steinhauer14,steinhauer16,nova19,kolobov21,steinhauer21}. In the present paper we focus on event horizons in BECs where excitations obey the Bogoliubov dispersion relation, but many of our results apply to event horizons in general.

Tidal bores (shock waves that travel up rivers due to a rising tide being funnelled into an V-shaped estuary) are a dramatic wave phenomenon that also involves hydrodynamic event horizons, as pointed out by Michael Berry in his studies of undular bores \cite{berry399_2007,berry505_2018}. His work emphasizes that the wave front corresponds to a \textit{caustic} where two waves coalesce, and this is the approach we adopt in this paper. The connection between event horizons and caustics is illuminating because caustics take on certain universal shapes described by catastrophe theory that are structurally stable against perturbations (and hence generic), and furthermore each catastrophe is dressed by a unique wave function. In the simplest case of two coalescing waves the caustic is a fold catastrophe and the wave function is the Airy function \cite{berry105_1981}. References \cite{berry399_2007,berry505_2018} show that the wave profile of an undular bore can indeed be expressed in terms of an integral which has the Airy function as its kernel.
 
In this paper we find that in the vicinity of a horizon in a BEC the waves satisfy a third order differential equation in position space which is similar to the one obeyed by undular bores. However, we generalize the treatment given in \cite{berry505_2018} to the case where the wave frequency $\omega$ is nonzero and show that this leads to an Airy-like wave function but with an additional logarithmic contribution $\omega \log k$ to the phase, giving a ``log-Airy function''. A logarithmic term has previously been shown to arise in aeroacoustic flows from jet engines \cite{howls18}, in certain path integrals in radio astronomy as studied by Feldbrugge et al from a wave catastrophe point of view (albeit one which differs from both our own and the standard approaches) \cite{feldbrugge2019,feldbrugge2020}, and also for gravitational black holes when dispersion is present \cite{coutant14}. The fact that a logarithmic phase singularity is a very general feature of waves in accelerated frames (such as near event horizons) has been highlighted by U. Leonhardt and collaborators  \cite{leonhardt02,kiss2004}, who first suggested a connection between horizons and wave catastrophes in a general sense, although a characterization in terms of the Thom-Arnold catastrophe theory which underlies standard caustics was not pursued. Links between caustics, Airy functions, and black hole physics have also been previously reported in more detail by Nardin et al in the context of water waves with subluminal dispersion \cite{nardin2009}. They pointed out the presence of a fold catastrophe bifurcation by taking a `dynamical systems' approach which is similar in spirit to ours. However, we will show in this paper that the logarithmic phase term connects the wave structure near the horizon to higher wave catastrophes than the fold.

 Indeed, the main motivation of the present paper is the observation that the coefficient multiplying the logarithmic term in the log-Airy function adds a second control parameter and thus the underlying catastrophe has some of the character of a cusp catastrophe, the next in Thom's hierarchy of catastrophes beyond the fold (which only has a single control parameter). This feature is supported by classical solutions of Hamilton's equations describing the motion of wavepackets which show that an event horizon gives rise to a broken pitchfork bifurcation in ($z,k$) phase space where a single solution bifurcates into three. The universal wave function dressing a cusp catastrophe is the Pearcey function, but unlike the Pearcey function the logarithm leads to mathematical complications in the evaluation of the integral representation of the solution, such as the need to introduce branch cuts and multiple Riemann sheets. We describe a systematic method for handling these complications using exponential coordinates \cite{howls18}. Along the way we shall identify instances of Stokes's phenomenon which occurs when an exponentially small wave appears behind a dominant wave and constitutes the ``quietly beating heart of asymptotics'' \cite{berry_trieste_2014}. In particular, we find that the horizon itself is the edge of a Stokes surface, giving some physical meaning here to this mathematical concept.

\section{Acoustic Event horizons}

\subsection{Bogoliubov dispersion relation}

Elementary excitations in a BEC obey the Bogoliubov dispersion relation \cite{pitaevskii03,pethick08}
    \begin{equation}
        \omega^2 =c^2 k^{2}\bigg(1+\frac{k^{2}}{4 k_{c}^{2}}\bigg)  \label{nonLinDisp}
    \end{equation} 
where  $c = \left(m\frac{\partial n}{\partial P} \right)^{-1/2}$ is the speed of sound expressed in terms of the mass $m$ of the atoms and the compressibility of the gas $\partial n / \partial P$ ($n$ is the number density and $P$ the pressure). $k_{c}=mc/\hbar$ is analogous to the Compton wavenumber (inverse of the healing length) \cite{leonhardt2003,leonhardt03b}, and provides the characteristic scale at which the Bogoliubov relation becomes dispersive: at small wavenumbers such that $k \ll k_{c}$ the relation reduces to the linear form $\omega=c k$ but for $k>k_{c}$ the relation curves upwards and hence is supersonic, or ``superluminal'' in the gravitational context. This implies that there are no true event horizons in BECs because there can be waves with arbitrarily large speed. However, these are energetically suppressed and, crucially, there is still a bifurcation when the flow speed exceeds the speed of sound so that  essentially the same phenomena arise as in fluids with subluminal dispersion relations \cite{Rousseaux2008,nardin2009,leonhardt2012,coutant14b}.
In a Lorentz invariant system the dispersion is purely linear, however this leads to divergences at the horizon known as the trans-Planckian problem which can be resolved by the introduction of dispersion \cite{unruh95,leonhardt2003,leonhardt03b,coutant14,jacobson91,brout95,corley96,corley98,Himemoto2000,schutzhold2005,agullo06,agullo07,coutant12,barbado11,isoard19}.

    \begin{figure}[t]
        \centering
	\includegraphics[width=12cm]{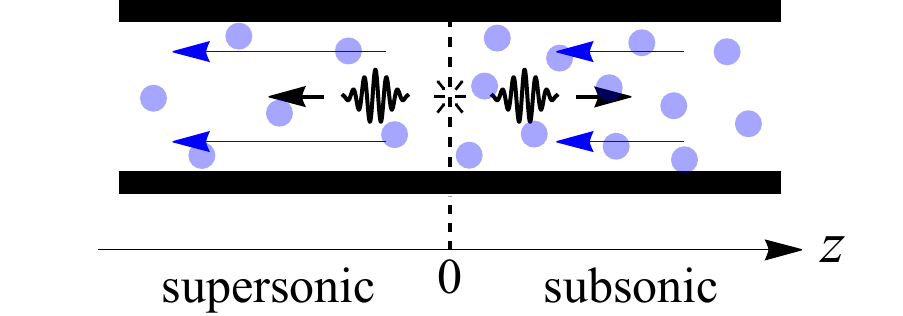}
        \captionsetup{width=1\linewidth}
        \caption{Schematic representation of a quasi-one dimensional acoustic black hole formed within a flowing fluid. The blue circles represent the constituent particles of the fluid, and the blue arrows indicate their direction and magnitude of flow. The vertical dashed line at $z=0$ indicates the position of an acoustic event horizon so that the flow is supersonic when $z<0$ (inside the black hole), and subsonic when $z>0$ (outside). The wavepackets represent Hawking radiation (pairs of back-to-back phonons) that propagate away from the horizon. One phonon escapes and moves rightwards, while the other gets swept leftwards into the black hole.} \label{constriction}
    \end{figure}
    
A schematic representation of a localized event horizon in a fluid is shown in figure \ref{constriction}. 
An early suggestion for experimentally realizing such a situation was to consider a BEC flowing around a ring that has a localized constriction such that the flow speed inside the constriction is forced to increase above the speed of sound in order to maintain the current and there is no buildup of atoms \cite{garay2000}. This set-up actually generates two horizons: a black hole horizon where the fluid enters the constriction and from which long wavelength phonons cannot escape and a white hole horizon on the other side where phonons cannot enter.  In the eventual experiments performed by the Steinhauer group a potential step (formed by a laser) is swept through a long thin cigar-shaped BEC and this effectively creates a waterfall over which the atoms flow \cite{lahav10}. We shall not attempt to model the details of these situations but instead content ourselves with the simplest theoretical model and consider a quasi-one dimensional BEC with a stationary velocity profile $u(z)$ that close to the horizon changes linearly in space \cite{leonhardt2012,coutant12}  
    \begin{equation}
        u(z)\approx -c+\kappa z 
        \label{flowVel}
    \end{equation}
where the velocity gradient $\kappa$ (analogus to the surface gravity of a graviational black hole) is positive and $c$ is the speed of sound at the horizon. This describes a flow from right to left whose magnitude is linearly increasing in the direction of the flow. At the point $z=0$ the flow speed is the same as the speed of sound and hence this is the location of the horizon: for $z>0$ the flow is subluminal and for $z<0$ it is superluminal (the speed of sound in a BEC made of a dilute gas of ultracold atoms is of the order of millimetres per second so this is easy to achieve).

The frequency that appears in Eq.\ (\ref{nonLinDisp}) is that in the fluid's rest frame, but an observer in the laboratory frame (where the horizon is fixed in space) will see a Doppler shifted frequency. A Galilean shift between the two frames gives the relation  $\omega' = \omega - u(z) k$ where here and from now on we use the unprimed notation to represent the laboratory frame. Thus, $\omega$ will be the frequency (energy) of the excitations seen from the laboratory frame. It is constant if the flow is time independent and must obey $\omega \geq 0$ in order to be physical. Conversely, $\omega'$ can be negative and this leads to an analogy to antiparticles.

\subsection{Nondispersive case: trans-Planckian singularity}

\begin{figure}[ht]
        \centering
        \captionsetup{width=1\linewidth}
            \begin{minipage}{0.5\textwidth}
                \includegraphics[width=7.4cm]{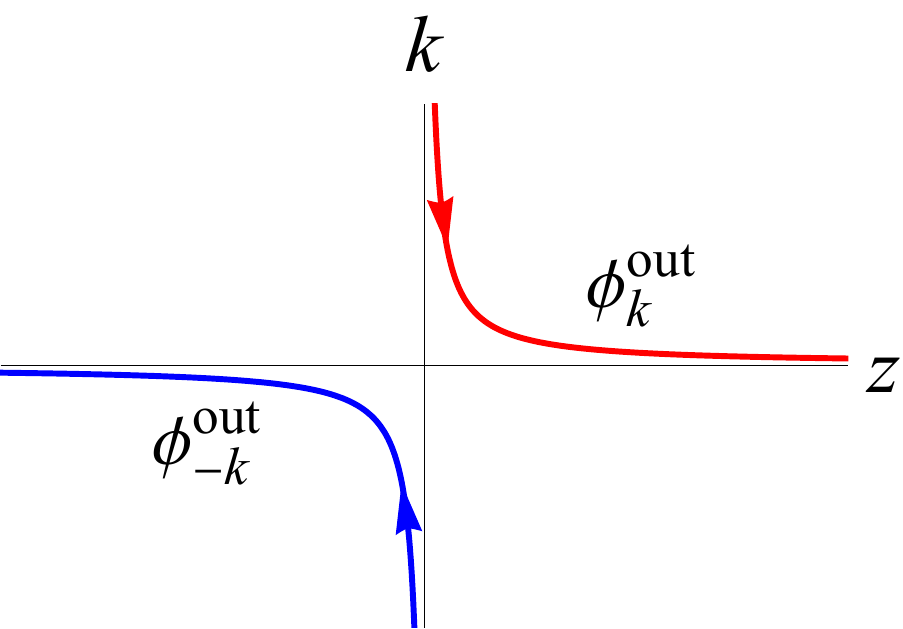} 
            \end{minipage}%
            \begin{minipage}{0.5\textwidth}
                \centering
                \includegraphics[width=7.4cm]{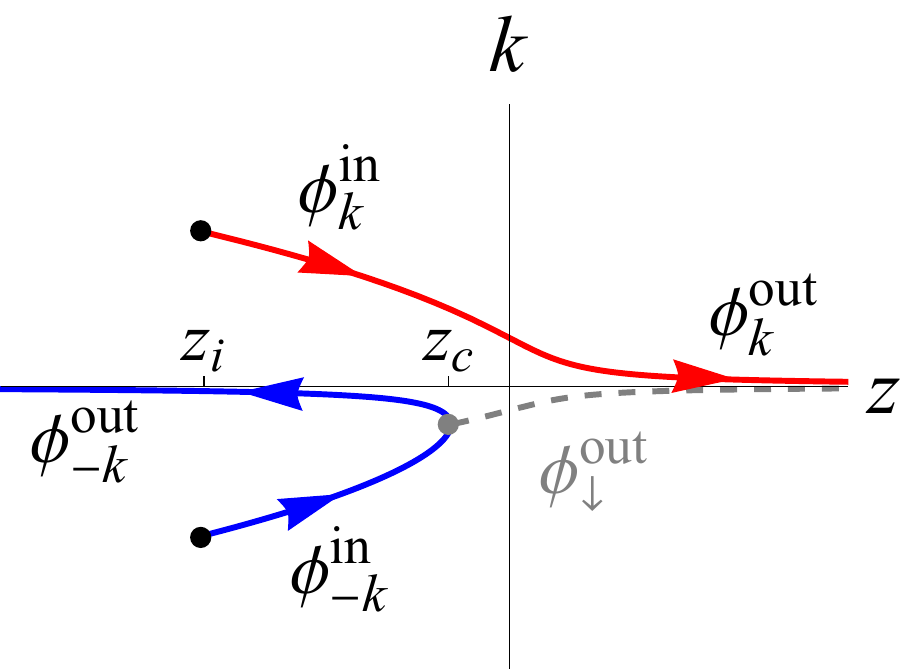}
            \end{minipage}
        \caption{\textbf{Left:} Solutions of Hamilton's equations [Eq.\ (\ref{hamEqs})] for linear dispersion [Eq.\ (\ref{linDisp})] with a finite value of $\omega>0$. The plot is in the $(z,k)$ phase space describing the position and wavenumber of wavepackets. The BEC is flowing from right to left: the right hand side of this figure is outside the black hole, and the left hand side is inside it. The red and blue branches $\phi^{\mathrm{out}}_{\pm k}$ are classical analogues of Hawking pairs that propagate away from the horizon towards $z= \pm \infty$. The wavenumbers diverge at the horizon. \textbf{Right:} Same as the leftmost plot but now for a nonlinear dispersion [Eq.\ (\ref{approxBogDisp})]. The resulting broken pitchfork bifurcation is the structure associated with a cusp catastrophe.  Both solutions start at the initial point $z_{i}$, and the critical point $z_{c}$ marks the turning point of the blue branch and is the location of a caustic where two real solutions merge to become two complex rays. Of these, only the decaying complex ray $\phi_{\downarrow}^{\mathrm{out}}$ is physical and is denoted by the gray dashed line.}\label{dispPlot}
    \end{figure}

We will first consider the long wavelength nondispersive regime where the nonlinear terms in the Bogoliubov dispersion relation can be ignored. The linearized dispersion relation in the laboratory frame is given by
    \begin{equation}
        \big(\omega-u(z) k\big)^{2} = c^{2}k^{2} \label{linDisp}.
    \end{equation}
In the limit where the flow varies very little over the wavelength of the produced sound waves, one can use a ray picture (geometrical acoustics). Treating $\omega(z,k)$ in Eq.\ (\ref{linDisp}) as a hamiltonian function, Hamilton's equations read
    \begin{equation}
        \frac{\mathrm{d}z}{\mathrm{d}t}=\frac{\partial \omega}{\partial k},\quad\quad\quad\quad \frac{\mathrm{d}k}{\mathrm{d}t}=-\frac{\partial \omega}{\partial z} \ . \label{hamEqs}
    \end{equation}
These rays describe the centre of mass motion $\langle z \rangle (t)$ and $\langle k \rangle (t)$  of wavepackets \cite{leonhardt2003,leonhardt03b}. Eqns.\ (\ref{hamEqs}) can be numerically integrated in time starting from an initial choice of position and wavenumber $(z_{i}, k_{i})$, and in the left image of figure \ref{dispPlot} we have plotted the trajectories for two different initial conditions, one just inside and one just outside the horizon, the idea being that these show the fate of spontaneous excitations starting near $z=0$. We see that in the phase space provided by the canonical variables $(z, k)$ the horizon behaves like a hyperbolic fixed point: rays move away from the horizon and tend asymptotically to $z= \pm \infty$ whilst the magnitude of their wavenumbers are reduced asymptotically to zero (red shifting). These two solutions can be interpreted as the classical manifestation of Hawking pair production where one excitation escapes the black hole and the other is trapped inside  \cite{leonhardt2012}. Indeed, both solutions have the same (positive) value of the lab frame frequency $\omega$ but have different signs of the fluid rest frame frequency $\omega'=\omega - u(z)k$ and hence can be considered to be a particle-antiparticle pair. If we instead run the time evolution backwards we find that these solutions of the linear model track back to the horizon where they develop wavenumbers of infinitely large magnitude $\pm k$. This is the trans-Planckian problem where the horizon seems to connect low energy physics to infinitely high energy physics. 

We have only included the positive roots of Eq.\ (\ref{linDisp}) in figure \ref{dispPlot}. The negative root also gives a physical solution, however, it describes waves that move with the flow which pass through the horizon relatively undisturbed and so will not be included here, not least because they are almost completely decoupled from the Hawking effect \cite{coutant14,coutant12}. However, including these other solutions is important if one wants total momentum $\sum_{i}k_{i}$ and total energy $\sum_{i} \omega-u(z)k_{i}$ conservation in the fluid rest frame (the laboratory frame frequency $\omega$ is automatically conserved by Hamilton's equations).

In the left plot of figure \ref{dispPlot} we have introduced the notation
$\phi_{k}^{\mathrm{out}}$ and $\phi_{-k}^{\mathrm{out}}$ to label the two solutions. This is adapted from the paper by Coutant, Parentani, and Finazzi (CPF) \cite{coutant12} upon which much of the present work is based. In particular, we use a wave scattering formalism  where ``out'' indicates that a wave is moving away from the horizon (in either direction), whereas the subscript $\pm k$ indicates whether the wave has a positive or negative wavenumber. In the dispersive problem we treat in the next section we will also have ``in'' waves that propagate towards the horizon. Finally, we note that within the linear theory the two sides of the horizon are disconnected: trajectories cannot cross the horizon.

\subsection{Wave theory in the nondispersive case: logarithmic phase singularity}

To obtain the effective wave equation that describes these waves we will follow Berry \cite{berry505_2018} and treat the frequency $\omega(z,k)$ as a hamiltonian which can be canonically quantized: $\omega(z,k) \rightarrow \hat{\omega}(\hat{z},\hat{k})$ where $z\rightarrow\hat{z}$ and $k\rightarrow\hat{k}=-\mathrm{i}\partial_{z}$. Using Eq.\ (\ref{linDisp}) we find that Schr\"{o}dinger's equation $\hat{\omega}\psi= \omega\psi$ for this system takes the form
    \begin{equation}
        -\mathrm{i} \kappa z\partial_{z}\psi(z,\omega)= \omega \psi(z,\omega) \label{linDiff}
    \end{equation}
        (the question of operator ordering in this equation is addressed in \cite{berry505_2018,coutant12}). This equation describes a stationary solution  $\psi(z,\omega,t)=\psi(z,\omega)\textrm{e}^{-\mathrm{i}\omega t}$ where
    \begin{equation}
        \psi(z,\omega)= A\textrm{e}^{\mathrm{i} \frac{\omega}{\kappa}\textrm{ln}(z)}, \label{linWaveFunc}
    \end{equation}
and $A$ is a constant. The logarithmic phase singularity at the horizon means the phase is undefined there and is a further manifestation of the trans-Planckian problem. Note that we have explicitly included the eigenvalue $\omega$ within the argument of $\psi$ to later make connections to catastrophe theory.

\subsection{Classical solutions in the dispersive case: broken pitchfork bifurcation}

In terms of the laboratory frequency $\omega$, the full Bogoliubov dispersion relation is
    \begin{equation}
        \omega-u(z) k =\pm ck\sqrt{1+\frac{k^{2}}{4 k_{c}^{2}}} \label{bogDisp}.
    \end{equation}
We will again keep only the positive root and furthermore will expand the right hand side and keep only the first nonlinear term. This is sufficient to capture the essential physics (a caustic where two or three waves coalesce depending on whether $\omega \neq 0$ or $\omega=0$, respectively). We therefore work with the cubic dispersion relation
    \begin{equation}
        \omega-u(z) k\approx c k+\frac{c k^{3}}{8 k_{c}^{2}} \label{approxBogDisp}.
    \end{equation}
The trajectories this generates via Hamilton's equations are plotted in the right image of figure \ref{dispPlot}. The initial conditions $(z_{i},k_{i})$ are indicated by black dots, and this time we choose the initial position $z_{i}$ for both trajectories to be inside the horizon as this allows a fuller picture of the dynamics: if the positive $k$ branch (red) were to start outside of the horizon it would continue to propagate to the right away from the horizon, whereas here we see that it passes through the horizon. Thus, the Bogoliubov dispersion relation allows the connection of the inside and outside of the black hole. In fact, comparison of both the left and right plots in figure \ref{dispPlot} shows that the effect of nonlinearity is to generate a broken pitchfork bifurcation such that for some values of $z$ inside the horizon there are three different possible values of $k$. The structure of the solutions means the divergence of the wavenumbers at the horizon is now eliminated but despite this change the Hawking process is robust to the introduction of dispersion because at large length scales we still have a pair of outgoing solutions $\phi_{\pm k}^{\mathrm{out}}$ and only the near-horizon behaviour is strongly modified \cite{unruh95,coutant14,jacobson91, brout95,corley96,corley98,Himemoto2000,schutzhold2005,agullo06,agullo07,coutant12,barbado11,isoard19}.

Like in the linear case, the negative $k$ solutions (blue) in the left plot of figure \ref{dispPlot} correspond to antiparticles because $\omega'<0$ when $k<0$. Their dynamics restricts them to be inside the horizon but this time they can travel in either direction. We recall that the motion of wavepackets is dictated by the group velocity $\mathrm{d} \omega / \mathrm{d} k = \mathrm{d}\omega'/\mathrm{d}k+u(z)$ and bearing in mind that in our setup $u(z) < 0$ this means $\mathrm{d}\omega/ \mathrm{d}k$ can be positive or negative irrespective of the sign of $k$, although for a large enough magnitude of $k$ it will always be positive due to the superluminality of the Bogoliubov dispersion.  
The critical point $z_{c}$ shown in the right plot of figure \ref{dispPlot} marks the place where $\phi_{-k}^{\mathrm{in}}$ turns around and becomes $\phi_{-k}^{\mathrm{out}}$ which is a caustic because turning points are places where two solutions coalesce. This coalescence ``interacts'' strongly with the third solution provided by the positive $k$ branch to give the broken pitchfork structure. This is the signature of a cusp-type catastrophe which has two control parameters: one determines where we are in the bifurcation (the $z$ coordinate) and the other determines the degree to which the pitchfork is broken (this will turn out to be $\omega$). Right at the cusp point itself, which occurs at $\omega=0$ and $z=0$ (see figure \ref{pearceyPlot}), we find an unbroken pitchfork where all three solutions coalesce. The caustic at $z_{c}$ also marks the point where two complex rays appear. One is physically reasonable since it corresponds to a decaying mode $\phi_{\downarrow}^{\mathrm{out}}$. The other corresponds to a growing mode $\phi_{\uparrow}^{\mathrm{out}}$ so we do not include it in our considerations  \cite{coutant12}.

\subsection{Length scale associated with quantum effects}

The wave function for the dispersionless case given in Eq.\ (\ref{linWaveFunc}) is scale free and has no intrinsic length scale to cut off the logarithmic divergence. However, the presence of $k_{c}$ in the dispersive case allows the introduction of a new length scale 
    \begin{equation}
        d=\left(\frac{c}{8 k_{c}^{2}\kappa}\right)^{1/3} 
    \end{equation} 
 which provides the characteristic distance over which the horizon becomes smeared out \cite{coutant14}. For the remainder of this work we will work in units where lengths are scaled by $d$, and times are scaled by $\kappa^{-1}$: 
    \begin{equation}
        \label{eq:zt_redefine}
        \frac{z}{d} \rightarrow z \quad , \quad  \kappa \, t \rightarrow t \quad , \quad k \, d \rightarrow k \quad , \quad \frac{\omega}{\kappa} \rightarrow \omega
    \end{equation}
 
The length $d$ has a quantum origin because it depends on the Compton wavelength $h/mc$ which sets the scale below which pair creation becomes important.
The role that $d$ plays in regulating  the logarithmic phase singularity  suggests that it is 2nd quantization of the classical wave theory that ultimately resolves our analogue trans-Planckian problem even though we do not explicitly 2nd quantize in this paper. In the classical theory of the pair production, the nonlinearity is sufficient to regulate it.

\subsection{Wave theory in the dispersive case: log-Airy function}

Canonically quantizing Eq.\ (\ref{approxBogDisp}) (in scaled units) yields a third-order linear ordinary differential equation in $z$-space 
	\begin{equation}
		i\big(\partial_{z}^{2}-z\big)\partial_{z}\psi(z,\omega)=\omega\psi(z,\omega) \label{thirdOrdDiff}.
	\end{equation}
This approximately describes $\psi(z,\omega)$ for both small $z$ and small $k$, although we keep in mind that although $k$ has been approximated to be small, it is not so small as to be equivalent to the linear dispersion approximation $\omega-u(z)k\approx ck$.
Ignoring the nonlinear dispersive effects equates to removing the $\partial_{z}^{2}$ operator within the brackets, and Eq.\ (\ref{thirdOrdDiff}) reduces to the linear case of Eq.\ (\ref{linDiff}) as expected. 

Although Eq.\ (\ref{thirdOrdDiff}) may be solved for $\psi(z,\omega)$ exactly in terms of a linear combination of three ${}_{1}F_{2}$ hypergeometric functions, it is more instructive to use a Fourier transform  to give an equivalent integral representation \cite{coutant14,coutant12}
    \begin{equation}
        \psi(z,\omega) \approx \frac{A}{2\pi}\int^{\infty}_{-\infty}\frac{1}{k}\textrm{e}^{-f(k,z,\omega)} \mathrm{d}k , \qquad
        f(k,z,\omega)=-\mathrm{i}\big[k^{3}/3+k z-\omega\textrm{ln}(k)\big], \label{loggedHawking}
    \end{equation}	
where $A$ is an arbitrary complex constant. Note that although Eq.\ (\ref{thirdOrdDiff}) is only valid for small $k$, we can extend the domain to include all $k\in (-\infty,\infty)$ by assuming that the integrand in Eq.\ (\ref{loggedHawking}) primarily contributes near $k=0$ and vanishes as $k$ becomes larger in both the $\pm k$ directions. With careful choice of cuts, this turns out to be true in our case.

The wave function in Eq.\ (\ref{loggedHawking}) is the primary object of interest in this work. We refer to it as the \textit{log-Airy function}
because the cubic polynomial in the exponential in the integrand is analogous to that of an Airy function  \cite{airy38,airy49} (see chapter 9 of \cite{DLMF} for the standard modern definition), which is the universal wave function that dresses a fold catastrophe \cite{thom75,arnold1975,poston78,gilmore81}, but differs due to the parametrically prefactored  additional logarithm in the phase. There is a temptation to combine the logarithm with the pole term to produce a complex order branch cut,  however, as we shall see, it proves analytically important to retain this term in the exponent. 

Eq.\ (\ref{loggedHawking}) has appeared before in a number of related contexts. The most closely related to our present treatment is the study of black hole radiation in Lorentz-violating theories in the papers by Coutant and Parentani \cite{coutant14} and CPF \cite{coutant12},
where the coefficient multiplying the logarithmic term is inversely proportional to the surface gravity which in turn sets the Hawking temperature. Similar logarithmic catastrophes have also been studied by Stone et al \cite{howls18,howls17} in the context of aeroacoustic noise, although Eq.\ (\ref{loggedHawking}) differs from \cite{howls18,howls17} in that it contains a $1/k$ pole term within the integrand, and its integration range is from $k\in(-\infty,\infty)$. An integral of the form of Eq.\ (\ref{loggedHawking}) also appears in Berry \cite{berry505_2018} in the context of tidal bores.  Berry works in a frame of reference that moves along with the bore such that the bore is stationary, which is equivalent to setting $\omega=0$.  This collapses Eq.\ (\ref{loggedHawking}) back to the integral studied by Boyd \cite {Boyd92} without the logarithm within the phase.

\section{The log-Airy function}

\subsection{Strategy}

Our goal now is to investigate the behaviour of Eq.\ (\ref{loggedHawking}) by characterizing its behaviour in the $(z,\omega)$ plane through a study of saddlepoint contributions and Stokes's phenomenon  that determine the physical ray and wave behaviour. The negative $k$ range of the integration leads to differences (and simplifications) from the work of Stone et al \cite{howls18,howls17}. In particular, the complex logarithm within the phase requires a choice of branch cut in order to define $\textrm{ln}(k)$ when $k<0$. Following CPF \cite{coutant14,coutant14b,coutant12,beck18}, there are two physically motivated choices of branch cut which can be made for our system in the complex $k$-plane: 
    \begin{enumerate}
        \item along the negative imaginary $k$-axis, leading to $\textrm{ln}(k)=\textrm{ln}\big(\vert k\vert\big)+i\pi$ when $k<0$,
        \item along the positive imaginary $k$-axis, leading to $\textrm{ln}(k)=\textrm{ln}\big(\vert k\vert\big)-i\pi$ when $k<0$.
    \end{enumerate} 
We will refer to (i) as the $+i\pi$ choice of branch cut, and (ii) as the $-i\pi$ choice.
Through large $\vert z\vert$ asymptotics of Eq.\ (\ref{loggedHawking}),  CPF determine that not all the waves scattered by the horizon can be obtained by a single choice of branch cut, and a description involving contributions from both is required. In fact, a complete solution of the third order differential equation Eq.\ (\ref{thirdOrdDiff}) requires three linearly independent global modes obtained by different choices of branch cuts and contours. Luckily one of the modes (the one associated with the exponentially growing wave  $\phi_{\uparrow}^{\mathrm{out}}$) is not needed for the Hawking problem and we will ignore it here. 

Our approach differs from CPF because we use exponential coordinates that allow us to focus on the near horizon behaviour in greater detail. In particular, CPF deform the contour of integration in the $k$-plane of Eq.\ (\ref{loggedHawking}) towards steepest descent paths over the saddles. However, some of the resulting deformed contours snag on the branch cuts and do not intersect saddles. In order to stay on the principal Riemann sheet, they make the approximation that $\omega$ is small and pull the logarithm within the phase out of their asymptotic approximation (see equation 57 in \cite{coutant12}). To deal with the snagged contours, they use an additional approximation referred to as the ``dominated convergence theorem'' which requires that $k_{c} \rightarrow \infty$ (so that the dispersion is approximated to be linear). These two separate approximations result in a partial loss of the near horizon wave structure, although it accurately describes the asymptotic behavior sufficiently far from the horizon.
By contrast, exponential coordinates allow us to treat multiple Riemann sheets in a transparent manner and we freely pass through the branch cut onto other sheets where we pick up additional saddlepoints in a straightforward way, without the need for additional asymptotic approximations. Furthermore, our method allows us to resolve caustics (coalescence of saddlepoints) and Stokes's phenomenon (sudden change of contour as new saddlepoints appear) which are the elemental processes that determine the structure close to the horizon.

We shall treat the lab frame frequency of the emitted quanta $\omega$ as a real parameter. For the system in question $\omega$ is small and positive (to observe the Hawking effect). The details of the calculation will be explained for this case, although a similar approach can be taken for both larger $\omega$ and $\omega<0$. 

\subsection{The large parameter, self-similar scaling properties and the classical limit}

Saddlepoint methods work well in the semiclassical regime where the phase oscillates rapidly (apart from at the saddlepoint itself). A large parameter $\lambda$ that performs the role of an inverse Planck's constant $\hbar^{-1}$ can be introduced into the log-Airy function  by changing variables in Eq.\ (\ref{loggedHawking}) from $k\rightarrow s$ where $k=\lambda^{1/3}s$, resulting in
    \begin{equation}
        \psi(z,\omega) = \frac{A}{2\pi} \textrm{e}^{-\mathrm{i}\frac{\omega}{3}\ln \lambda}\int^{\infty}_{-\infty}\frac{1}{s}\textrm{e}^{-\lambda f(s,z,\omega)} \mathrm{d}s , \qquad 
        f(s,z,\omega)=-\mathrm{i}\big[s^{3}/3+s z/\lambda^{2/3}-(\omega/\lambda)\textrm{ln}(s)\big]. \label{loggedHawkingscaled}
    \end{equation}	
Like in the cases of the standard diffraction integrals (Airy, Pearcey, etc.), this procedure gives redefined control parameters, in this case $z/\lambda^{2/3}$ and $\omega/\lambda$. Therefore, in addition to the redefinitions given in Eq.\ (\ref{eq:zt_redefine}), we henceforth make the further redefinition
    \begin{equation}
       \frac{z}{\lambda^{2/3}} \rightarrow z \quad , \quad \frac{\omega}{\lambda} \rightarrow \omega  \ .  \label{eq:redefinition2}
    \end{equation}
The powers ($2/3,1$) to which $\lambda$ is raised are known as Berry indices and determine how the fringe spacing in the wave function evolves in the directions specified by the control parameters as $\lambda$ is changed \cite{berry58_1977}.  
The index $2/3$ matches that of the Airy function, but the unity power for $\omega$ does not have a counterpart in the standard diffraction integrals, see table 36.6.1 in \cite{DLMF} (the Pearcey function has the indices $3/4$ and $1/2$). Another difference to the standard diffraction integrals is that the scaling does not change the overall magnitude of $\psi(z,\omega)$ since the new factor $e^{-i(\omega/3)\ln \lambda}$ outside the integral is a pure phase term. This is different to the Airy case for which the amplitude diverges as $\lambda^{1/6}$, where the exponent 1/6 is the Arnol'd singularity index \cite{DLMF,arnold1975}. For the Pearcey function it is $1/4$. Thus, the classical limit $\lambda \rightarrow \infty$ leads to an infinitely rapidly varying phase of the wave function but does not lead to an infinite amplitude as it does for standard caustics. 

What physical parameter should we choose for $\lambda$? The natural choice is the quantum length scale $d$ defined in Eq.\ (\ref{eq:zt_redefine}). More precisely, we choose the dimensionless ratio
    \begin{equation}
        \lambda=\frac{d_{0}}{d},
        \label{eq:lambda_defn}
    \end{equation}
where $d_{0}$ is an arbitrary reference length scale. The classical limit in our problem is therefore $d_{0}/d \rightarrow \infty$ where pair creation occurs only at vanishingly small length scales and the spectrum is linear to infinitely large values of the wavenumber. Recalling that $z$ in Eq.\ (\ref{loggedHawkingscaled}) is already scaled by $d$, we find that the redefined coordinate in Eq.\ (\ref{eq:redefinition2}) grows as $d^{-1/3}$. In other words, the fringe spacing in physical space shrinks, as expected.

Our approach to the large parameter needed for the saddlepoint analysis is different to that employed in other works such as that by CPF where their parameter (equation 34 in \cite{coutant12}) is spatially dependent and vanishes at the horizon. By contrast, our $\lambda$ is spatially constant which is important because we seek to resolve the saddlepoint structure even at the horizon.

\subsection{$\omega=0$ case and pole contribution}

To make a link with previous work, we first study the case when $\omega=0$, which is related to a zero-energy ``soft mode". In that case the logarithmic term in Eq.\ (\ref{loggedHawkingscaled}) vanishes and it reverts to an integral which appeared in Berry \cite{berry505_2018} in the context of tidal bores. The integrand then becomes effectively an Airy-function exponent with a pole and may be evaluated as an integral of the Airy function itself and was studied by Boyd \cite{Boyd92}.  

\begin{figure}[t]
        \centering
            \begin{minipage}{0.9\textwidth}
                \includegraphics[width=15cm]{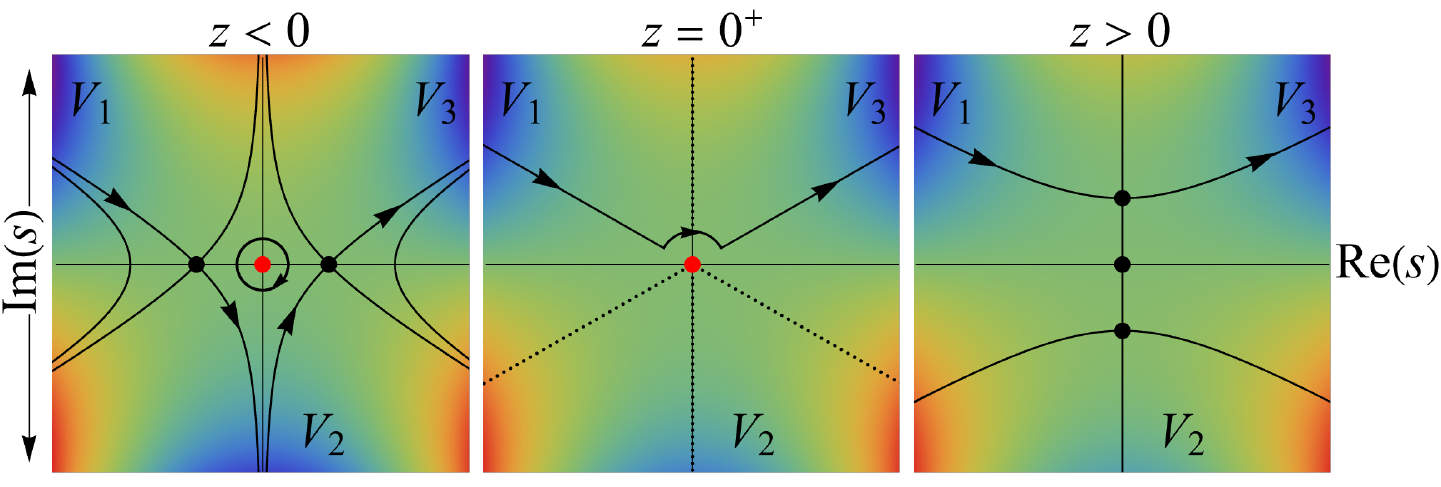}
            \end{minipage}%
            \begin{minipage}{.17\textwidth}
                \centering
                \includegraphics[width=0.34cm]{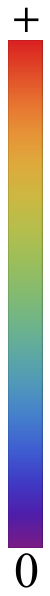}
            \end{minipage}
        \captionsetup{width=1\linewidth}
        \caption{Steepest descent plots for the log-Airy function in the $s$-plane when $\omega=0$. Each panel is for a different value of $z$: $z<0$ (inside), $z=0^{+}$ (on), and $z>0$ (outside) the black hole horizon. The dots denote the saddlepoints at $s=\pm\sqrt{z}$ and the pole at $s=0$. The steepest descent paths must run between valleys $V_{1,2,3}$ (blue) as $|s|\rightarrow+\infty$ where the Re$[-\lambda f(s,z,w)]<0, \lambda>0$ ($\arg(k)=5\pi/6$ and $\pi/6$) so that the integral converges. Specifically, paths must start in $V_{1}$ and end in $V_{3}$, but may take an excursion to and from the intermediate $V_2$ valley. Red regions denote regions as $|s|\rightarrow+\infty$ where Re$[-\lambda f(s,z,w)]>0$. The steepest paths pass over the pole at $s=0$ for $z\le 0$ forcing a residue contribution. The red dots denotes the values of $z$ where the pole contributes, black where it does not. The case of $\omega=0$ shown in this figure is special because the horizon and the caustic coincide (at $z=0$).}
        \label{omega=0summaryfig}
    \end{figure}

This integral contains three asymptotic contributions, two from the saddlepoints and one from the pole, see figure \ref{omega=0summaryfig}. There is but a single caustic point at $z=0$, where the two saddles coalesce with the pole at $s=0$.  As $z$ runs from $z<0$ to $z>0$, the steepest descent contours in the $s$-plane will deform and reconnect as is the case for the Airy function. As the system passes through $z=0$ the contours unavoidably cross the pole and so generate an additional residue contribution, essential for the step function of the initial data in the case of the bore. The overall result is the well known bore form of a step function modulated by an Airy function.

By contrast, we shall see for $\omega\ne 0$ below that although the waveform at fixed $z$ mimics that of a bore, the rise in the overall magnitude for $z<0$ is not due to the pole.  Rather an equivalent analysis shows that steepest paths in the $s$-plane do not ever cross the pole at $s=0$. They deform around it, but never generate a residue from it. The modulated step function appearance of the exact result can be seen to be generated from pairs of saddlepoint contributions, where one of the saddles comes from adjacent Riemann sheets.

\subsection{Similarity to cusp catastrophe}

\begin{figure}[t]
        \centering
        \captionsetup{width=1\linewidth}
        \includegraphics[width=15cm]{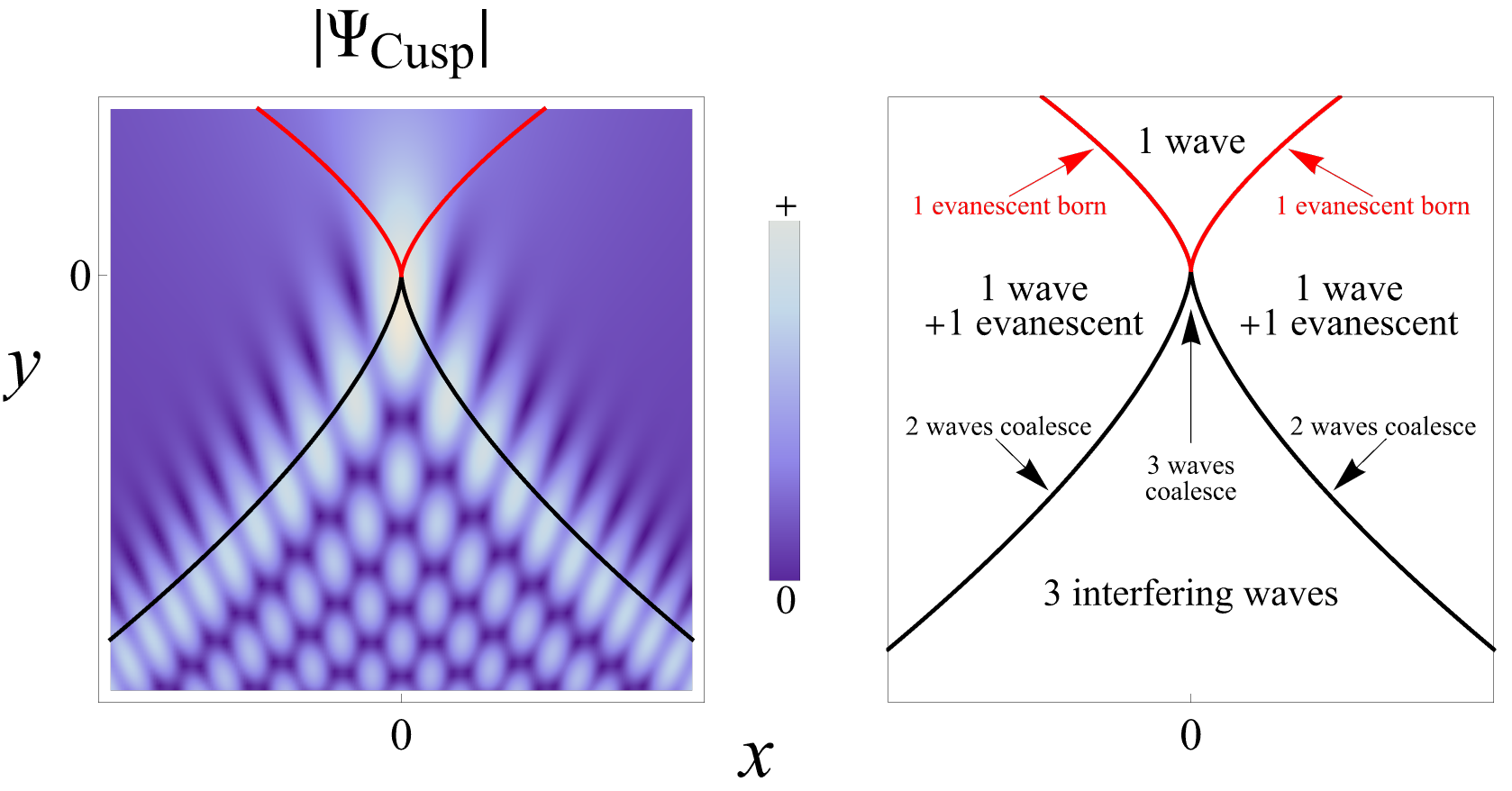}
        \caption{\textbf{Left:} Modulus of the Pearcey function for $\lambda=1$ as defined by Eq.\ (\ref{pearceyFunc}). The black curve indicates the cusp caustic where two waves coalesce, except at the very tip (origin) where three waves coalesce. The red curve indicates the Stokes set \cite{wright80}, where an evanescent wave is born. \textbf{Right:} The number of waves (saddles) in each region relative to the cusp caustic and the Stokes set. The notation ``wave'' is short for ``real wave'' (i.e.\ non-evanescent).} \label{pearceyPlot}
    \end{figure}

We now study the case of $\omega \ne 0$. The location of the saddlepoints on the principal sheet of the $k$-plane is given by  $\partial_s f(s, z,\omega)=0$.  For fixed $\omega\ne 0$ the locations of the saddles $s_j$ are therefore given by
    \begin{equation}
        s_j^2+z-\omega/s_j=0, \qquad \Rightarrow \qquad s_j^3+zs_j-\omega=0, \qquad j=1,2,3.	\label{loggedsaddles}
    \end{equation}
Hence the presence of the logarithm increases the number of saddlepoints by one to three, rather than the two underpinning the Airy function. Thus, we expect the analytical skeleton of the log-Airy function Eq.\ (\ref{loggedHawkingscaled}) to be more akin to that of the next most complex function in the catastrophe theory hierarchy, the cusp, whose waveform is given by the Pearcey function \cite{DLMF,thom75,poston78,gilmore81} 
    \begin{gather}
        \Psi_{\textrm{Cusp}}(y,x;\lambda)=\sqrt{\frac{\lambda}{2\pi}}\int_{-\infty}^{\infty}\textrm{e}^{i \lambda(s^{4}/4+y s^{2}/2+x s)}ds	\label{pearceyFunc}, \\[0.3cm]
        \textrm{Pe}(y,x)=\Psi_{\textrm{Cusp}}(y,x;1) \ . \nonumber
    \end{gather}
The diffraction pattern generated by this integral is depicted in the left hand panel of figure \ref{pearceyPlot} and features a two dimensional set of fringes generated by three wave interference below a cusp-shaped caustic (black curve). The highest intensity fringes are close to the caustic which is the place on which the saddles of Eq.\ (\ref{pearceyFunc}) coalesce in pairs except right at the tip of the cusp where all three saddles coalesce. Asymptotic expansions about the saddlepoints diverge on the caustic but the exact waveform given by Eq.\ (\ref{pearceyFunc}) is smooth: interference between the waves resolves the ray singularity on the caustic. The overall wave behavior of the Pearcey function relative to the caustic (depicted by the black curve) is shown in the right panel of figure \ref{pearceyPlot}. The red curve gives the position of the Stokes set for the Pearcey function (as first found by F.\ J.\ Wright \cite{wright80}). 

The ray limit of the Pearcey function, which provides the scaffold upon which the waves are hung, is obtained from the saddlepoints of the exponent of the integrand in Eq.\ (\ref{pearceyFunc}) which satisfy $s_{j}^3+ys_{j} +x=0$. This equation is identical to Eq.\ (\ref{loggedsaddles}) for the log-Airy function with $(x,y)$ playing the roles of $(-\omega,z)$. This implies that,
at least as far as real rays are concerned, the ray structure for the log-Airy function is identical to that of the cusp. 

The position of the caustic for the log-Airy function is given by the condition for coalescence of two or more saddles by the simultaneous satisfaction of $\partial_s f(s, z,\omega)=0$ and $\partial^2s f(s, z,\omega)=0$.  Eliminating $s$ from these two equations yields
    \begin{equation}
        27\omega^{2}+4z^{3}=0.	\label{loggedCaust}
    \end{equation}
Since the real saddles of the log-Airy function obey the same equation as those of the Pearcey function the classical caustic structure  for the two functions is also identical. In fact, the local similarity to the next highest form in the catastrophe hierarchy of functions will be true for all such cuspoid diffraction integrals perturbed by logarithms in their exponent \cite{DLMF}. 

    \begin{figure}[h]
        \centering
        \captionsetup{width=1\linewidth}
            \begin{minipage}{0.9\textwidth}
                \includegraphics[width=14cm]{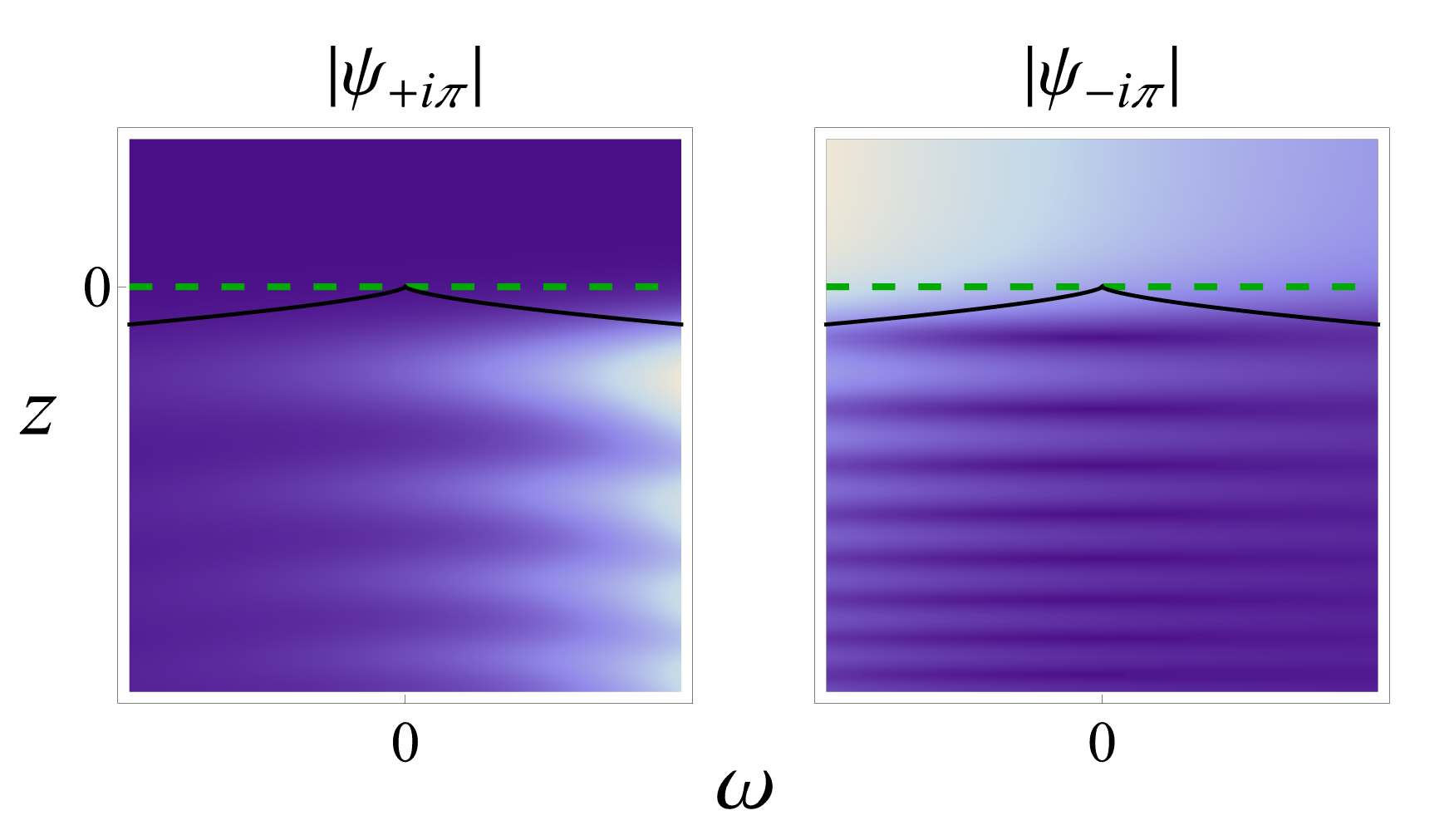} 
            \end{minipage}%
            \begin{minipage}{.05\textwidth}
                \centering
                \includegraphics[width=0.34cm]{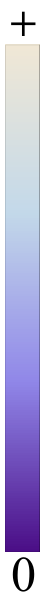}
            \end{minipage}\\[0.3cm]
            \begin{minipage}{.5\textwidth}
                \includegraphics[width=7.8cm]{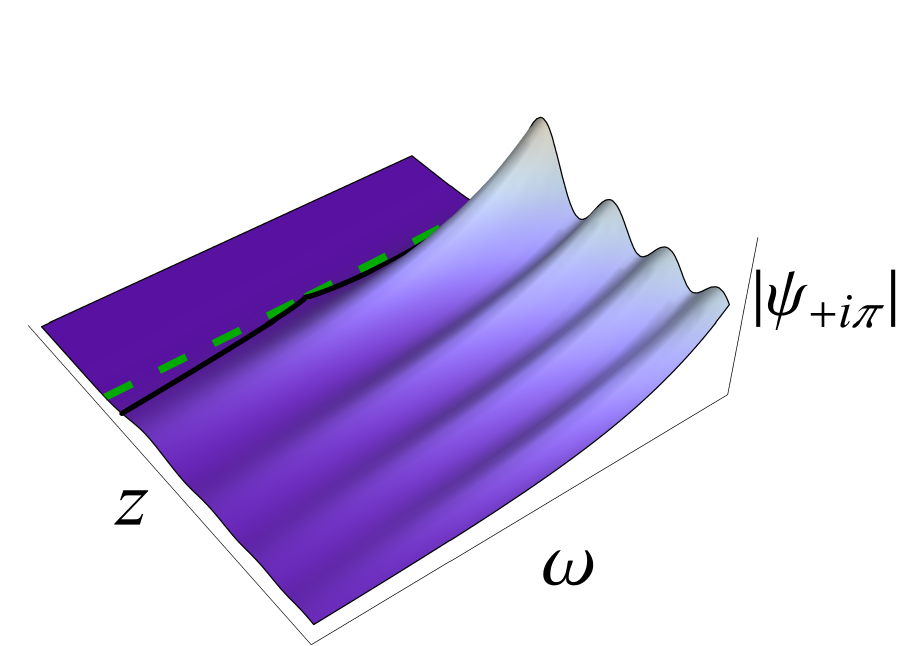} 
            \end{minipage}%
            \begin{minipage}{.5\textwidth}
                \includegraphics[width=7.8cm]{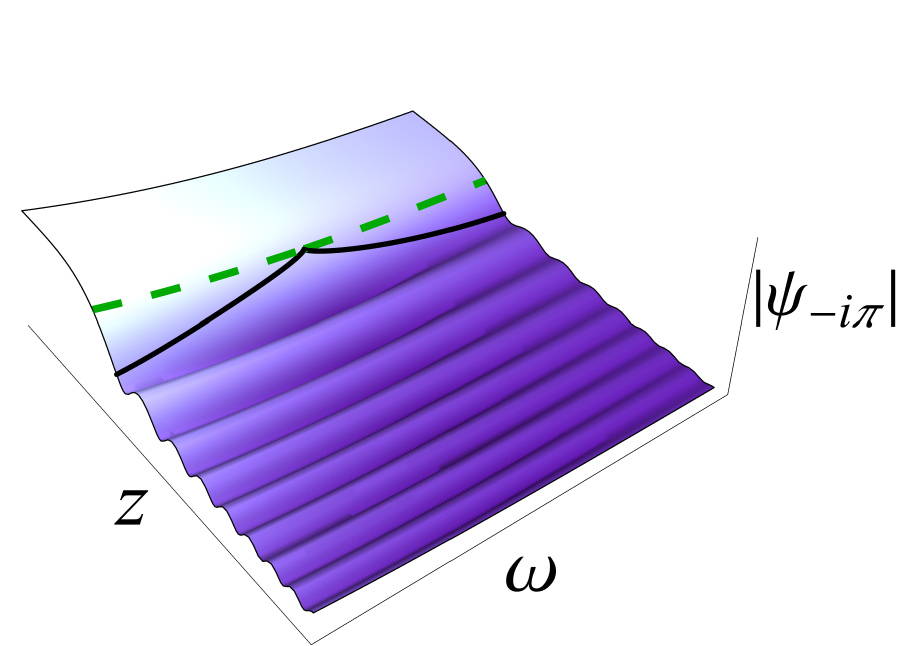} 
            \end{minipage}
        \caption{Density (\textbf{upper row}) and surface (\textbf{lower row})  plots of the log-Airy function defined by Eq.\ (\ref{loggedHawkingscaled}) (with $\lambda=1$) for the $+i\pi$ (\textbf{left column}) and $-i\pi$ (\textbf{right column}) choices of branch cut. The white areas are where the intensity begins to diverge as $\omega$ becomes sufficiently positive or negative, depending on the choice of branch cut. The black curves in each image represent the caustic, given by Eq.\ (\ref{loggedCaust}), and the green dashed lines give the location of the event horizon at $z=0$, where the flow velocity $u=-c$. The scale in the $z$ direction has been compressed in relation to that of figure \ref{pearceyPlot} so as to show the oscillations below the caustic.}\label{fig:loggedHawking}
    \end{figure}

Despite the identical caustic structure for the log-Airy and Pearcey functions, their waveforms differ significantly. 
It is a relatively straightforward numerical calculation to evaluate the log-Airy integral in Eq.\ (\ref{loggedHawkingscaled}) and the results for $\lambda=1$ for both choices of branch cut are displayed in figure \ref{fig:loggedHawking}. The two dimensional fringe pattern of the Pearcey function inside the cusp caustic and its damped nature outside are not replicated for the log-Airy function. Rather, for values of $z$ below the caustic the log-Airy function has one dimensional fringes. For the same range of $\omega$ these are more pronounced for the choice of $+i\pi$ cut. For the same choice of cut, the magnitude of Eq.\ (\ref{loggedHawkingscaled}) grows indefinitely as $\omega\rightarrow +\infty$ and decays when $\omega\rightarrow -\infty$. The opposite is true for the $-{\rm i}\pi$ cut: the amplitude grows for $\omega\rightarrow -\infty$ and decays when $\omega\rightarrow \infty$. We now seek to explain this behaviour using a steepest descent analysis.

\subsection{Steepest descent contour diagrams}

    \begin{figure}[t]
        \centering
        \includegraphics[width=10cm]{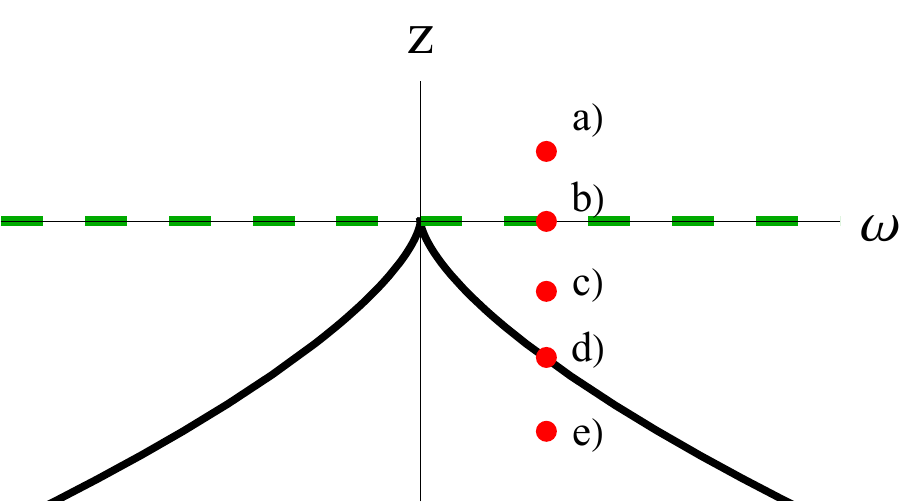} 
        \captionsetup{width=1\linewidth}
        \caption{Points of interest in the $(z,\omega)$ plane relative to the caustic (solid black curve) and horizon (green dashed line). Each point resides in a distinct region and will be studied in detail in figures \ref{ipibranchsummary} and \ref{mipibranchsummary}. Points a) resides upstream of the horizon outside of the black hole, point b) resides on the horizon, and the remaining points c) through e) reside within the black hole: points c) are between the horizon and the caustic, point d) is on the caustic, and points e) are downstream of the caustic. Although we will only study these specific points, the respective regions they occupy also display equivalent behavior for $\omega>0$. For $\omega<0$ the dominance of contributing saddles is simply the opposite of the corresponding $\omega>0$ regions, but we only focus on $\omega>0$ in this paper.}\label{saddRegions}
    \end{figure}

We perform a careful steepest descent analysis of Eq.\ (\ref{loggedHawkingscaled}), extending the work of CPF \cite{coutant12} to study particular regions of interest relative to the \textit{horizon and caustic}. Due to the fact that the horizon and the caustic do not coincide except at $\omega=0$, there is effectively a broadened horizon (gap between horizon and caustic) on the length scale of $d$ which grows in width as $\delta z \propto \omega^{2/3}$ \cite{leonhardt2003,coutant12}. This broadening is seen in figure \ref{dispPlot} as the gap between the turning point located at $z_{c}$ and the horizon at $z=0$, and also as the region between the solid black and dashed green curves in figure \ref{fig:loggedHawking}.

We identify five distinct points a) through e) as indicated in figure \ref{saddRegions}, and will apply our method at these points for each choice of branch cut, starting with the $+i\pi$ cut, although our method can just as well be applied for the $-i\pi$ choice (as will later be shown) and to the $\omega<0$ half plane. Each steepest descent contour diagram is created in the $s$ integration plane for a particular set of control parameters $(z,\omega)$.
To ensure convergence of the integral the real $s$ contour must be deformed into paths of steepest ascent/descent starting and ending in asymptotic valleys at infinity $V_{1}$ and $V_{3}$ respectively where 
$${\rm Re}[\lambda f(s,z,\omega)]\rightarrow +\infty, \qquad |s|\rightarrow +\infty,$$
passing through saddles $s_j$, $j=1,2,3$ of the phase and satisfying  
$${\rm Im}\left\{\lambda[f(s,z,\omega)-f(s_j,z,\omega)]\right\}=0.$$
The steepest descent contours may take excursions to and from intermediate valleys, for example, $V_2$ or its copies on different Riemann sheets. The arguments of $s$ that determine the asymptotic valleys will depend on which choice of cut is taken for the logarithm, but copies on different sheets will always have values of $\arg(s)$ separated by $2\pi$. 

Different subsets of saddles $s_j$ can contribute at different values of $(z,\omega)$.  This is as a result of the topology of the steepest descent paths changing at a Stokes's line in the $(z,\omega)$ plane. The steepest paths emerging from two (or more) saddles $i\ne j$ connect in the complex $s$-plane when 
$${\rm Im}\left\{\lambda[f(s_i,z,\omega)-f(s_j,z,\omega)]\right\}=0.$$
If the number of saddlepoints contributing to the asymptotic expansion of the integral changes as a Stokes line is crossed this is termed a Stokes's phenomenon, see for example \cite{berry181_1989}.

\subsection{$+i\pi$ branch cut}

\begin{figure}[t]
        \centering
        \captionsetup{width=1\linewidth}
            \begin{minipage}{.43\textwidth}
                \includegraphics[width=7cm]{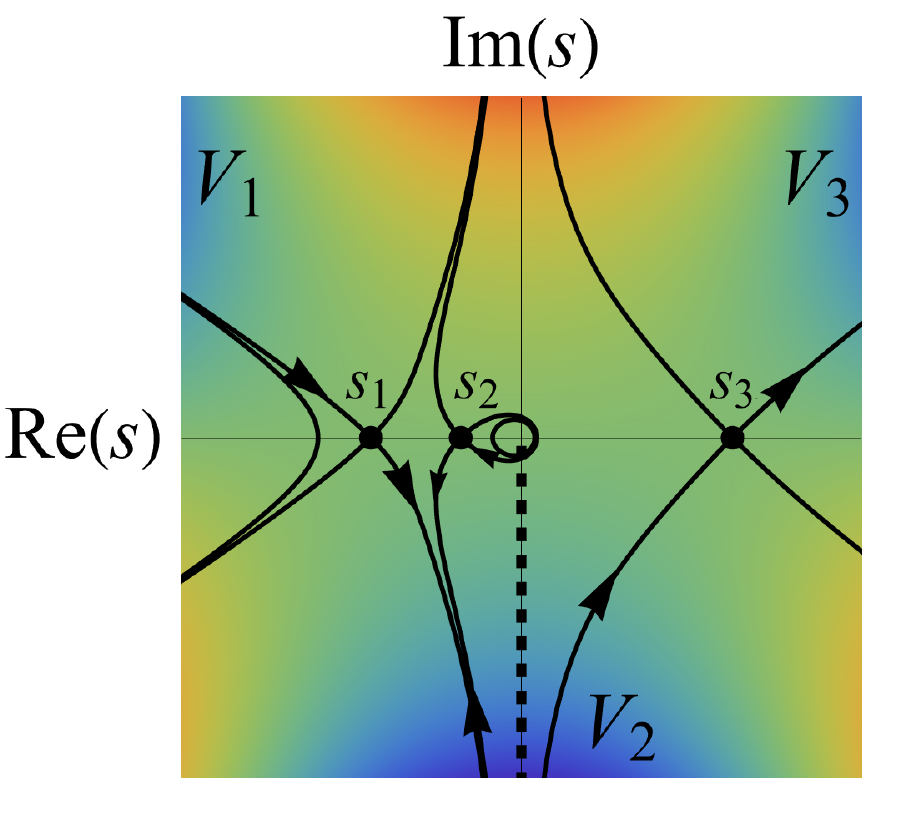} 
            \end{minipage}%
            \begin{minipage}{0.1\textwidth}
                \centering
                \includegraphics[width=0.34cm]{graphics/ampBar2.pdf}
            \end{minipage}%
            \begin{minipage}{.5\textwidth}
                \includegraphics[width=7cm]{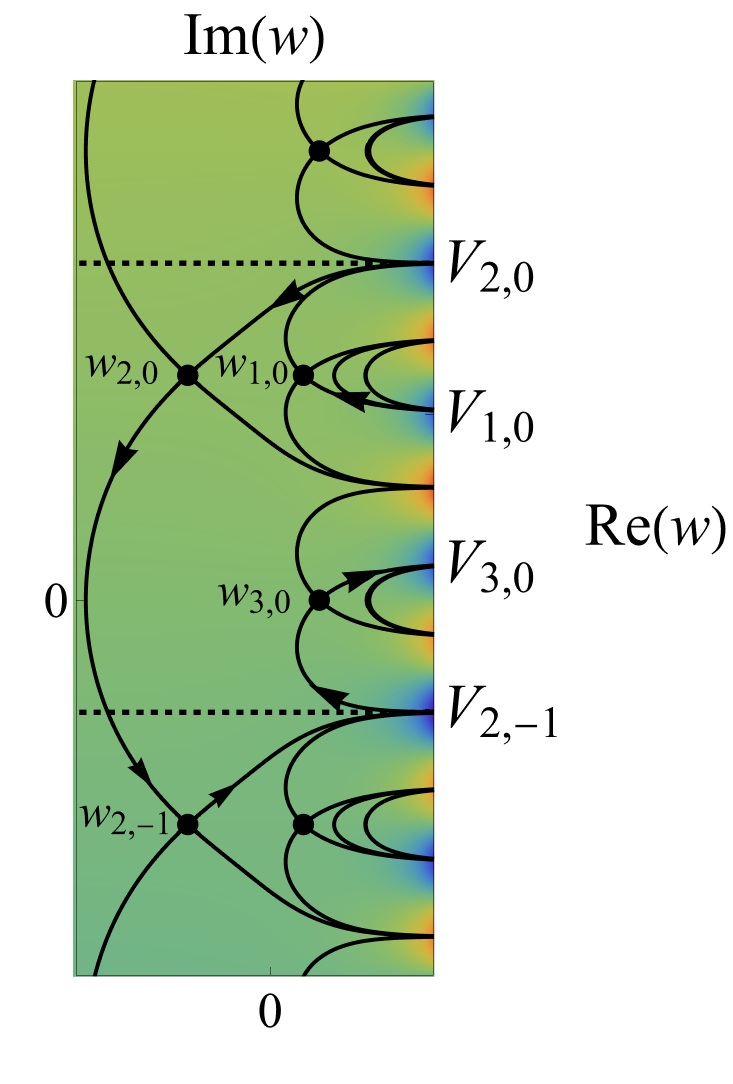} 
            \end{minipage}\\[0.3cm]
        \caption{\textbf{Left:} The steepest paths (solid black lines) in the original $s$ integration plane for a choice of control parameters ($z,\omega$) corresponding to all points e) in figure \ref{saddRegions}, together with the choice of $+i\pi$ branch cut (dashed black line). The black dots denote the saddlepoints $s_{j}$ of the phase, given by Eq.\ (\ref{loggedsaddles}), and the arrows indicate the contributing (converging) steepest descent contours. \textbf{Right:} Equivalent steepest descent contour diagram in the transformed $w$-space. The horizontal dashed lines denote the mappings of the (now no longer) $+i\pi$ branch cut, and the black dots are the unfolded saddlepoints $w_{j,n}$. In both the left and right panels the larger arrows indicate contributing contours on the principal sheet, and the smaller arrows indicate those contributing on adjacent sheets. It is clear that it is easier to follow the contributing contour in $w$-space than in $s$-space.}\label{loggedContourW1a}
    \end{figure}

We shall first focus on the region characterized by points e) in figure \ref{saddRegions}. The steepest descent contour diagram of Eq.\ (\ref{loggedHawkingscaled}) at these points is given by the left hand plot in figure \ref{loggedContourW1a}.
The presence of the cut complicates progress in the $s$-plane. The dashed black line indicates the $+i\pi$ branch cut, and at first glance there seems to be no obvious choice of contour (solid black line) starting in $V_{1}$, passing through any of the saddles (black points), and ending in $V_{3}$. 

CPF \cite{coutant12} proceeded by effectively removing the logarithmic term from the phase (which requires $\omega$ to be small) for the purposes of saddlepoint analysis such that the analysis reduces to that of the Airy function modified by the logarithmic branch cut. They allowed their deformed steepest descent contours to snag on the branch cut and expanded asymptotically around that loop contour. This loop contour is not along a path of steepest descent, but is approximately evaluated in terms of a complex gamma function by the ``dominated convergence theorem" ($k_{c}\rightarrow\infty$), which to zeroth order ignores nonlinear dispersive effects. 

Here, we instead continue to follow the steepest paths, even as they encounter the branch cut and flow onto adjacent Riemann sheets. 
The result is a calculation that then relies just on simple expansions around saddlepoints including, where required, on the non-principal sheet. This has a potential for an easier physical interpretation than the loop contour around the cut, and additionally does not neglect any near horizon behavior since the nonlinearity in the dispersion is taken into account.  
We discuss the relative differences of the approach of CPF \cite{coutant12} and this paper further in the discussion below.

To that end, we follow Stone et al  \cite{howls18}, and apply the exponential transformation
    \begin{equation}
        s=e^w
    \end{equation}
to the integrand of Eq.\ (\ref{loggedHawkingscaled}) to give
	\begin{equation}
		\psi(z,\omega) = \frac{A}{2\pi}\int^{V_{3,0}}_{V_{1,0}}\textrm{e}^{-\lambda f(w,z,\omega)} dw , \qquad 
	f(w,z,\omega)=-\mathrm{i}\big(e^{3w}/3+ze^w-\omega w\big) \ . \label{loggedwplane}
	\end{equation}	
This unfolds the infinite number of Riemann sheets arising from the logarithm in the complex $s$-space, whose boundaries are defined by the choice of $+i\pi$ branch cut, into a single complex $w$-space. It also removes the need to consider the pole contribution in the original integral for $\omega\ne 0$ (when $\omega =0$, this transformation generates a third saddlepoint at $\omega\rightarrow -\infty$ equivalent to the pole).

Each sheet is mapped to a horizontally stacked semi-infinite strip of height $2\pi i$ in the $w$-plane, starting with this choice of branch cut at $w=+i\pi$.  
As a consequence, the $w$-plane contains an infinite number of periodically stacked valleys $V_{j,n}$ and saddlepoints 
\begin{equation}
w_{j,n}=w_{j,0}+2n\pi i, \label{wSaddles}
\end{equation}
where $j=1,2,3$, and $n\in \mathbb{Z}$ denotes the Riemann sheet on which the valleys or saddles sit ($n=0$ being the principal sheet for the $+i\pi$ choice of branch cut). There is  a basis of 3 saddlepoints on the principal sheet, with copies equally spaced out at a complex distance $2\pi i$ on each mapped sheet.

The $w$-plane representation is shown in the right hand plot in figure \ref{loggedContourW1a}. The solid black lines in the $w$-plane denote the images of the steepest descent contours. The horizontal black dashed lines denote the mappings of the (now no longer) $+i\pi$ branch cut. The blue regions denote the asymptotic valleys of convergence, and the red regions are the asymptotic hills where the integration along a contour would diverge. In $w$-space the periodically repeating valleys $V_{j,n}$ all lie along the positive Re$(w)$ side as $w\rightarrow \infty$. The arrows in both plots of figure \ref{loggedContourW1a} indicate the direction of travel along the now continuously deformed steepest path staring from $V_{1,0}$ in the principal sheet, passing through a subset of the saddles $w_{j,n}$, on different Riemann sheets if needs be, before ending back at $V_{3,0}$ back on the principal sheet.

In $w$-space starting at $V_{1,0}$ within the principal sheet, the contour intersects the first saddle $w_{1,0}$ and runs off into the $V_{2,0}$ valley ``above it", along the bottom of the branch cut at $w=+i\pi$. The deformed contour then re-emerges from $V_{2,0}$ intersects a second saddle $w_{2,0}$, before leaving the principal sheet to encounter $w_{2,-1}$ which is a copy of the saddle $w_{2,0}$ at a point $2\pi i$ vertically below on the next lowest sheet.  The contour then turns by a right angle (indicating Stokes's phenomenon) before passing into $V_{2,-1}$ which is a copy of $V_{2,0}$ on the lower side of border with the principal sheet. The last component of the contour re-emerges from $V_{2,-1}$ and passes through saddle $w_{3,0}$ before finally running into the asymptotic valley $V_{3,0}$ on the principal sheet. The contours are considerably easier to follow in $w$-space than $s$-space, and it is seen that all asymptotic contributions to the integral arise from saddlepoints (or the pole outside the exponential), rather than loops around branch cuts, allowing for a (local) application of catastrophe theory (the topological theory underlying the coalesence of stationary points) to understand the properties of the integral.

\subsection{Asymptotic contributions}

In terms of asymptotic contributions from saddlepoints in the original $s$ and transformed $w$-plane, we have the correspondence:
    \begin{equation}
        w_{1,0} \longleftrightarrow s_1, \qquad w_{2,0}\longleftrightarrow s_2, \qquad w_{2,-1}\longleftrightarrow s_{2,-1}, \qquad w_{3,0}\longleftrightarrow s_3,
    \end{equation}
where $s_{2,-1}$ is the image of saddle $s_2$ on the next lowest Riemann sheet in the $s$-plane.
The asymptotic contribution from the expansion about each saddlepoint $w_{j,n}$ along the doubly infinite steepest decent contour that passes through it takes the form \cite{BerryHowls}
    \begin{equation}
        \psi^{(j,n)}(z,\omega)= \sum_{r=0}^{N_{j,n}-1} \psi^{(j,n)}_{r}(z,\omega)\sim (-1)^q\frac{e^{-\lambda f_{j,n}}}{\sqrt{\lambda}}\sum_{r=0}^{N_{j,n}-1}{\frac{T^{(j,n)}_r(z,\omega)}{\lambda^r}},  \label{asympexp}
    \end{equation}
where, with  $f_{j,n}=f(w_{j,n},z,\omega)$,
    \begin{equation}
        T^{(j,n)}_r(z,\omega)=\frac{\Gamma(r+1/2)}{2\pi {\rm i}}\oint_{w_{j,n}}\frac{dw \  g(w)}{(f(w,z,\omega)-f_{j,n})^{r+1/2}}\label{termint}.
    \end{equation}
$N_{j,n}$ is the number of terms taken in the asymptotic series expansion for the corresponding $w_{j,n}$ saddlepoint, $\Gamma$ denotes the gamma function, $g(w)=1/(2\pi)$ which does not actually depend on $w$ and is constant in our case, and $q=0$ or $1$, depending on the direction of traversal of the contour relative to the computed $T^{(j,n)}_r$. When the expansion is undertaken over a semi-infinite contour, as occurs at a Stokes's phenomenon, additional terms at half powers of the large asymptotic parameter are required, see \cite{Howls92,Bennetetal}.

The terms in the expansion may be computed via the residue integral representation of Eq.\ (\ref{termint}) or via the Lagrange reversion method \cite{Dingle}. The first couple of terms in a doubly infinite contour expansion over $w_{j,n}$ are:
    \begin{equation}
         T^{(j,n)}_0(z,\omega)=\sqrt{\frac{2\pi}{f''_{j,n}}}g(w_{j,n}) = \frac{e^{-{\rm i}\pi/4}}{\sqrt{2\pi{\rm i}(3e^{3w_{j,n}}+z \mathrm{e}^{w_{j,n}})}}, \label{zeroTerm} 
    \end{equation}
    \begin{eqnarray}
         T^{(j,n)}_1(z,\omega)&=& \left. \frac{1}{12f''^{7/2}}\sqrt{\frac{\pi}{2}}\left(12f''^2g''+5gf'''^2-3f''(4g'f'''+gf'''')\right)\right\rvert_{w_{j,n}} \nonumber \\ &=&
        \frac{e^{-3{\rm i}\pi/4}(81e^{4w_{j,n}}+z^2)}{12\sqrt{2\pi}(3e^{2w_{j,n}}+z)^{7/2}}.
    \end{eqnarray}
The expressions for $f_{j,n}$ are algebraically complicated and will not be written out here but they, and hence the associated saddlepoint expansions, are valid for a general range of $\omega$. However, for the purposes of comparison with \cite{coutant12}, we only need their analytical form in the small $\omega$ limit (this is also the regime where the coupling between the positive and negative wavenumber solutions that gives the Hawking effect is strongest).

In order to demonstrate our method we focus our attention on points e) which lie below the caustic in figure \ref{saddRegions}, although it can and will be later applied to the remaining points (it could also be applied for $\omega<0$). Points e) correspond to saddlepoint diagrams equivalent to that of figure \ref{loggedContourW1a}. At such points, the principal sheet contributes three real (in $s$-space) saddles, describing three wave interference for $\omega>0$ inside of the horizon and below the caustic, like in the Pearcey function. However, here the outer two saddles in the left-hand plot of figure \ref{loggedContourW1a} give contributions that dominate the third, leading to the Airy-like interference pattern observed in the $\omega>0$ region of the $\psi_{+i\pi}$ plots in figure \ref{fig:loggedHawking}. Furthermore, this approach also reveals the presence of a contribution from a \textit{fourth saddle}, $w_{2,-1}$.
In this way we find that for the $+i\pi$ branch cut at points e), the formal asymptotic expansion $\psi^{(e)}_{+i\pi}(z,\omega)$ takes the form
    \begin{equation}
        \psi^{(e)}_{+i\pi}(z,\omega)\sim \psi^{(1,0)}(z,\omega)+\psi^{(2,0)}(z,\omega)+\frac{1}{2}\psi^{(2,-1)}(z,\omega)+\psi^{(3,0)}(z,\omega). \label{e-sum}
    \end{equation}
 The contribution from $w_{2,-1}$ is exponentially smaller than that from $w_{2,0}$ by a factor $e^{-2\lambda\pi\omega}$ ($\omega>0$). Such real exponential factors are associated with pair production \cite{coutant14,coutant14b,coutant12}.
The factor of 1/2 is due to the presence of the Stokes's phenomenon which may be inferred from the contour intersecting this saddle making a sudden sharp ``dog-leg" turn as it encounters the saddle (see the right hand panel in figure \ref{loggedContourW1a}), a characteristic signature of a Stokes's phenomenon.

Due to the fact that the $T^{(j,n)}_r(z,\omega)$ effectively only depend on derivatives of $f(w,z,\omega)$ at $w_{j,n}$, it is easy to see that for each $r$,
    \begin{equation}
        T^{(2,0)}_r(z,\omega)=T^{(2,-1)}_r(z,\omega).
    \end{equation}
Taking into account the relative sense of the traversal of the contours over $w_{2,0}$ and $w_{2,-1}$ we find that Eq.\ (\ref{e-sum}) simplifies at leading order to
    \begin{equation}
       \psi^{(e)}_{+i\pi}(z,\omega)\sim  \frac{1}{\sqrt{\lambda}}\left\{ e^{-\lambda f^{(1,0)}}T_0^{(1,0)}+\left(1+\frac{1}{2}e^{-2\lambda\pi\omega}\right)e^{-\lambda f^{(2,0)}}T_0^{(2,0)}+e^{-\lambda f^{(3,0)}}T_0^{(3,0)}\right\}. \label{e-sum2}
    \end{equation}
From Eqns.\ (\ref{eq:zt_redefine}) and (\ref{eq:redefinition2}) we see that in the notation of \cite{coutant12}, their $\omega/\kappa$ factors are equivalent to our $\lambda \omega$ factors. In order to compare against the results of CPF we observe that we need to consider the small $\omega$ regime: taking the  limit $\omega \rightarrow 0^{+}$ of $f_{j,n}$ and $T_0^{(j,n)}$, we find that the factors that contribute to Eq.\ (\ref{e-sum2}) can be written as
    \begin{equation}
      e^{-\lambda f^{(1,0)}}T_0^{(1,0)}\sim  \frac{e^{+\tfrac{2}{3}{\rm i}\lambda|z|^{3/2}}|z|^{-i\lambda \omega/2}e^{\lambda\pi\omega}e^{+3{\rm i}\pi/4}}{2\sqrt{\pi}|z|^{3/4}},\label{T1}
    \end{equation}

    \begin{equation}
      e^{-\lambda f^{(2,0)}}T_0^{(2,0)}\sim  \frac{e^{{\rm i}\lambda \omega}e^{-{\rm i}\lambda\omega \log{\omega}}|z|^{i\lambda \omega}e^{{\rm i}\pi/4}e^{\lambda \pi \omega}}{\sqrt{2\pi\omega}},\label{T2}
    \end{equation}

    \begin{equation}
      e^{-\lambda f^{(3,0)}} T_0^{(3,0)} \sim\frac{e^{-\tfrac{2}{3}{\rm i}\lambda|z|^{3/2}}|z|^{-i\lambda \omega/2}e^{-3{\rm i}\pi/4}}{2\sqrt{\pi}|z|^{3/4}} \ . \label{T3}
    \end{equation}
    
Up to overall prefactors (see also the next paragraph), Eqns.\ (\ref{T1}) and (\ref{T3}) are consistent with the results given in equations 62 and 63 of \cite{coutant12} respectively, with the identification of CPF's $\Delta(z)=|z|^{3/2}$. The removal in \cite{coutant12} of the log term from the exponent before undertaking a steepest descent approach assumes both $\vert z \vert \gg 1$ and $\vert z \vert \gg \omega^{2/3}$ \cite{coutant14,coutant12}, and 
is consistent with the small $\omega$ approximation made here. However, our retention of the logarithm in the exponent increases the range of validity of the results in $\omega$ (albeit at the expense of algebraic complexity). A consequence of removing the logarithm from the saddlepoint exponent in \cite{coutant12} is that although this gives an accurate asymptotic approximation for large enough $z$, it does not capture the presence of a separate horizon and cusp caustic.

As we now explain, our Eq.\ (\ref{T2}), multiplied by the $(1+e^{-2\lambda \pi\omega}/2)$ factor from Eq.\ (\ref{e-sum2}), is equivalent to CPF's more complicated looking third contribution which is given by equation 64 in \cite{coutant12}. In that paper equation 64 does not come from a saddlepoint contribution but instead from a loop contour around the cut along -i$\mathbb{R}$ in the $s$-plane (the left image in their figure 4). The authors then apply the dominated convergence theorem which requires that $k_{c} \rightarrow \infty$ (equivalent to our $\lambda\rightarrow\infty$) and so to lowest order ignores the cubic term in the integral exponent. After notational translations, equation 64 in \cite{coutant12} contains terms involving $\sinh (\lambda \omega \pi) \Gamma(-{\rm i}\lambda\omega)$. Their result expands on the Stokes line of the gamma function according to equation 3.4 of \cite{ParisWood} for $z=-{\rm i}\lambda \omega$, with $\lambda \omega>0$ as:

    \begin{eqnarray}
       \Gamma(z)&\sim & \sqrt{2\pi}z^{z-1/2}e^{-z}\left\{1+\frac{1}{12z}+{\mathcal O}\left(\frac{1}{z^2}\right)\right\}(1-e^{-2\pi {\rm i} z})^{-1/2} \nonumber \\
       &\sim & \sqrt{2\pi}z^{z-1/2}e^{-z}\left\{1+\frac{1}{12z}+{\mathcal O}\left(\frac{1}{z^2}\right)\right\}(1+\frac{1}{2}e^{-2\pi {\rm i} z} +\dots). \label{ParisWoodEqn}
    \end{eqnarray}

Hence for large $\lambda$ the leading order result of equation 64 in \cite{coutant12} yields Eq.\ (\ref{T2}). The additional subdominant contribution in the second line of Eq.\ (\ref{ParisWoodEqn}), $+(1/2)e^{-2\pi {\rm i} z}=+(1/2)e^{-2\lambda\pi\omega}$, corresponds to the contribution of the fourth saddle $w_{2,-1}$ which, since $f^{(2,0)}-f^{(2,-1)}=2\pi{\rm i}$, combines to give the prefactor $(1+e^{-2\lambda \pi\omega}/2)$ in the overall $w_{2,0}$ term in Eq.\ (\ref{e-sum2}). From this, it can be seen that the pure-saddlepoint approach not only incorporates the cubic terms in the exponent but also avoids the need for complex gamma functions (and the complexity of the correct analytical representation of them on their Stokes lines) whether for small or more general values of $\omega$.

Physically speaking, and for values of $z$ below the cusp, the contributions $w_{1,0}$ [Eq.\ (\ref{T1})] and $w_{3,0}$ [Eq.\ (\ref{T3})] correspond to the two waves $\phi_{-k}^{\rm in}$ and $\phi_k^{\rm in}$ respectively, in the right plot of figure \ref{dispPlot} (these waves are responsible for the Airy-like interference in the wave plots shown in figure \ref{fig:loggedHawking}). The combination of $w_{2,0}$ and $w_{2,-1}$ [Eq.\ (\ref{T2}) multiplied by the $(1+1/2e^{-2\lambda\pi\omega})$ factor from Eq.\ (\ref{e-sum2})] generates the $\phi_{-k}^{\rm out}$ wave in the same region (responsible for the step function which is modulated by the Airy-like interference). The direction of travel of these three waves can be confirmed by realizing that the spatially dependent parts of the phases in Eqns.\ (\ref{T1})--(\ref{T3}) are given by the WKB result $\int_{0}^{z}k \, \mathrm{d}z'$, so that $k(z)$ can be obtained by differentiation, and the group velocity found from $v_{\mathrm{gp}}=(\mathrm{d}k/\mathrm{d}\omega)^{-1}$.  For the waves in Eqns.\ (\ref{T1}) and (\ref{T3}) this gives $v_{\mathrm{gp}}=2 \kappa d \vert z \vert/\lambda$ confirming that they are right moving. For the wave in Eq.\ (\ref{T2}) it is
$v_{\mathrm{gp}}= - \kappa d \vert z \vert/\lambda$ showing that it is left moving. Interestingly, the group velocity does not depend on the `Airy factors' $\pm (2/3) \mathrm{i} \lambda \vert z \vert^{3/2}$ as these do not depend on $\omega$.

\bigskip

\bigskip

\subsection{$+i\pi$ branch cut contributions}

\begin{figure}[t]
        \centering
        \captionsetup{width=1\linewidth}
        \includegraphics[width=15.75cm]{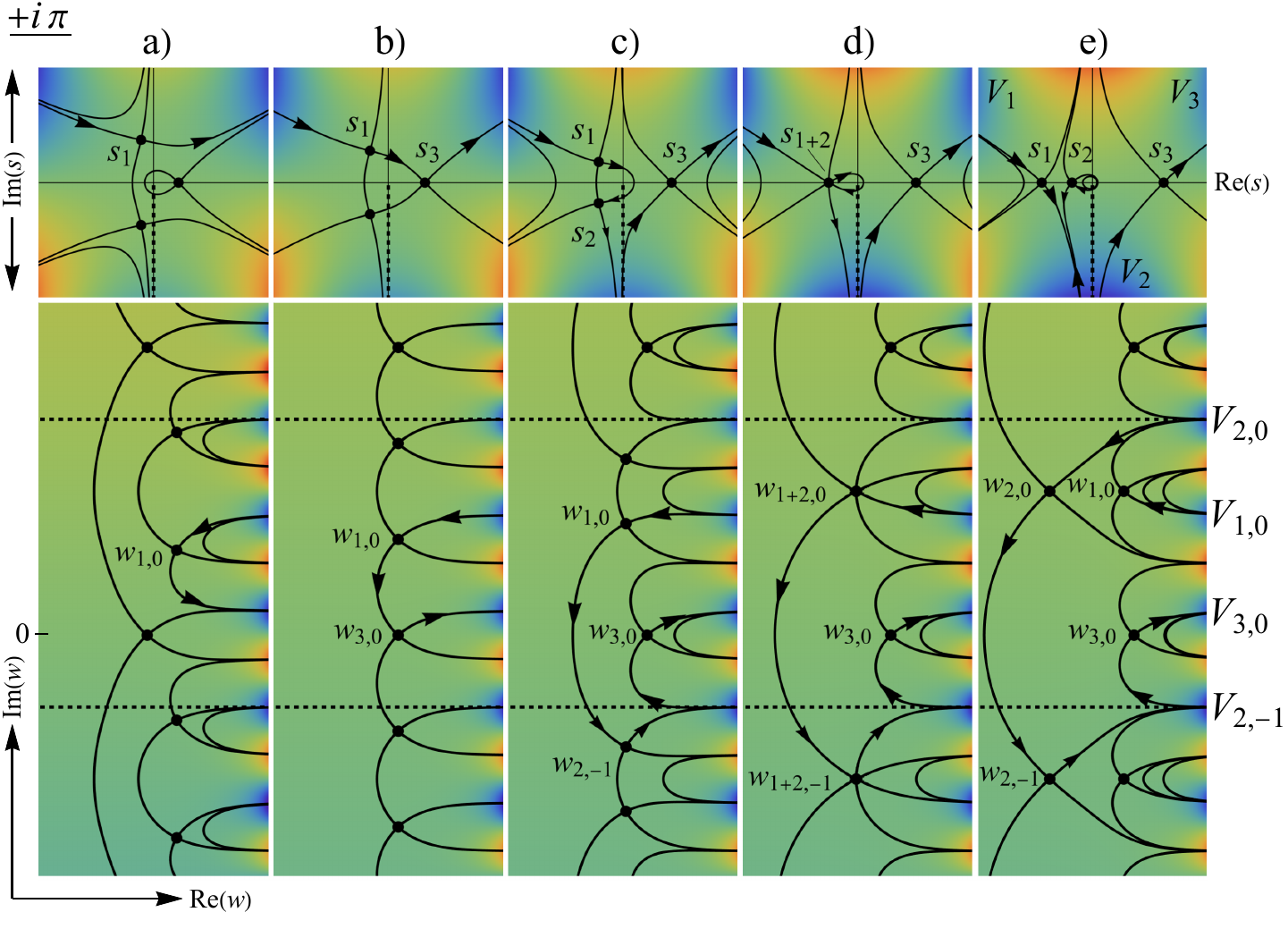} 
        \caption{Summary of the saddlepoint contributions at points a) through e) from figure \ref{saddRegions}, for the $+i\pi$ branch cut. The upper plots are the complex $s$-space contour diagrams, while the lower plots are the corresponding $w$-space ones. We use the notation $s_{i+j}$ $(i\neq j)$ to denote when multiple saddlepoints $s_{i}$ and $s_{j}$ coalesce (which happens at a caustic) and become equal to one another. We also adopt this notation for the $w$ saddles.} \label{ipibranchsummary}
    \end{figure}

We can proceed in the same way for each point in the $(z,\omega>0)$ half-plane. The qualitative steepest paths and the associated contributing saddles at these locations are displayed in figure \ref{ipibranchsummary}.
    
Along the caustic at point d), as expected, two of the saddles from e) (figure \ref{loggedContourW1a} or the rightmost column in figure \ref{ipibranchsummary}) have coalesced into one, $w_{1+2,0}$, and contribute whilst simultaneously undergoing a Stokes's phenomenon with an exponentially subdominant pair of coalesced copies $w_{1+2,-1}$, together with a simple real saddle $w_{3,0}$ with a purely imaginary phase. 

For values of $z$ that lie between the caustic and the horizon, corresponding to points c), there is one real saddle $w_{3,0}$ and one complex saddle $w_{1,0}$ undergoing a continuous Stokes's phenomenon with the contributing subdominant saddle $w_{2,-1}$. The latter saddle lies outside of the principle Riemann sheet, and all together the contributing saddles yield $\psi^{(c)}_{+i\pi}$. By a continuous Stokes's phenomenon we mean it occurs for a range of $z$ as opposed to at a single value of $z$.

At the horizon $z=0$, which is point b), $w_{1,0}$ now undergoes an instantaneous Stokes's phenomenon (i.e.\ only at the single point $z=0$) inside the principle sheet with the subdominant contributing saddle $w_{3,0}$. Finally, at points a) we see the real saddle no longer contributes and only a single complex saddle $w_{1,0}$ remains (corresponding to $\phi_{\downarrow}^{\textrm{out}}$ in the right plot of figure \ref{dispPlot}).  

Consideration of the steepest contours in figure \ref{ipibranchsummary} shows that there is a change in the number of contributions across the event horizon at $z=0$ ($w_{1,0}$ contributes on both sides, but $w_{3,0}$ only contributes for $z<0$). Hence the event horizon is a Stokes line for real $\omega$.   

 This is consistent with WKB analysis in a single complex dimension, where Stokes lines sprout from turning points, of which the caustic here is a higher dimensional version. The connection between certain types of black holes, horizons, pair-production, and Stokes's phenomenon has also been studied in various other contexts \cite{anderssonhowls,Dumlu_2020,Dumlu_2010,Hashiba_2022}. In our case the Stokes surface differs: it intersects real 2-parameter space in a line corresponding with the horizon.

\bigskip

\subsection{$-i\pi$ branch cut contributions}

    \begin{figure}[tbp]
        \centering
        \captionsetup{width=1\linewidth}
        \includegraphics[width=15.75cm]{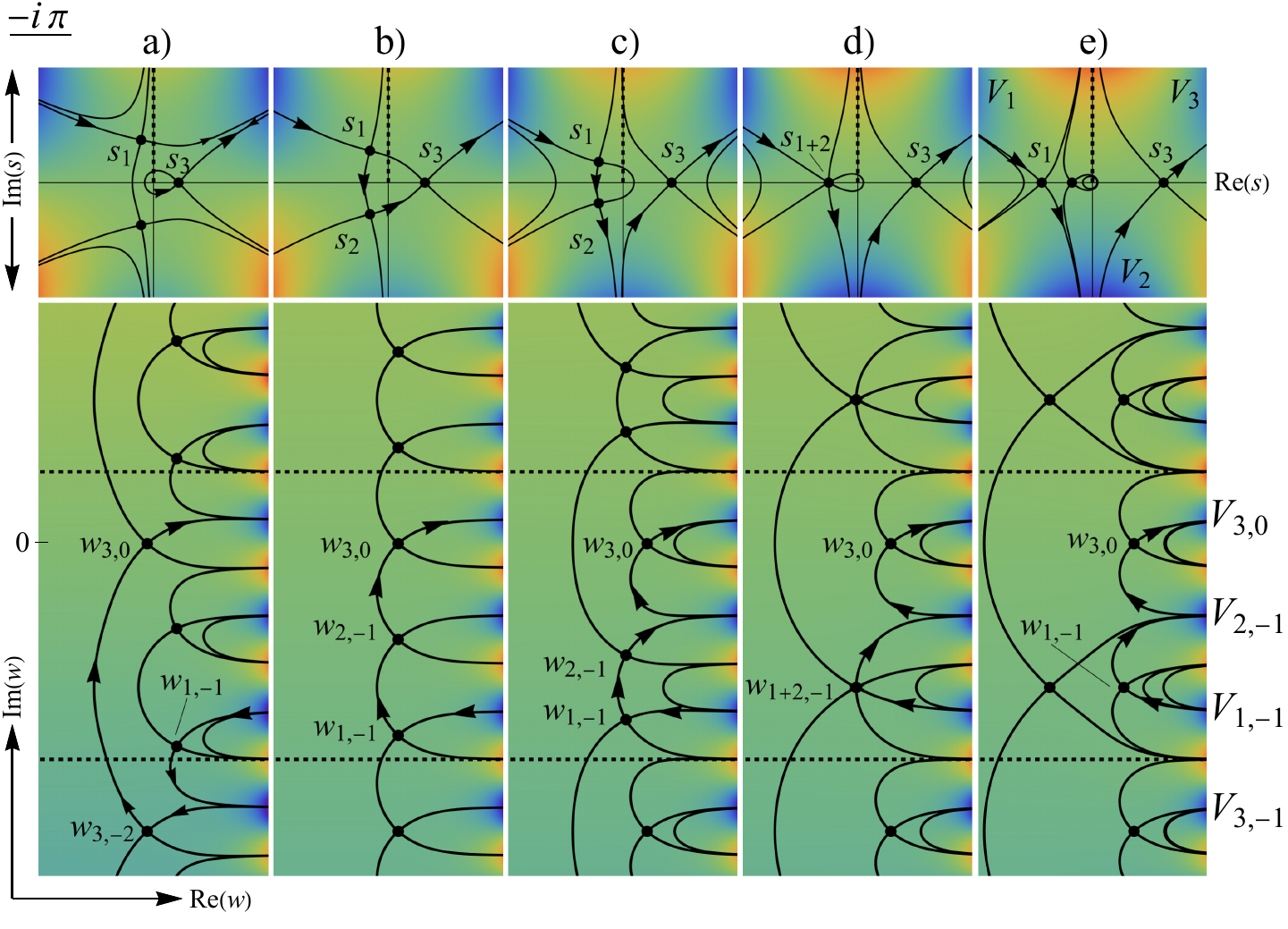} 
        \caption{Summary of the saddlepoint contributions at points a) through e) from figure \ref{saddRegions}, but this time for the $-i\pi$ branch cut. The upper plots are the complex $s$-space contour diagrams, while the lower plots are the corresponding $w$-space ones. The contours and saddles are the same as those shown in figure \ref{ipibranchsummary} for the $+i\pi$ choice of branch cut, but the Riemann sheets in $w$-space have all shifted down by $i\pi$ in comparison. This forces the initial and final valleys  ($V_{1,-1}$ and $V_{3,0}$, respectively) to differ from those in figure \ref{ipibranchsummary}, and thus different saddles contribute. For consistency we have kept the same labelling of saddles as in figure \ref{ipibranchsummary}.} \label{mipibranchsummary}
    \end{figure}

We can proceed in the same way for the  $-i\pi$ choice of branch cut. For both $s$- and $w$-space, the contours and saddles in figure \ref{mipibranchsummary} are exactly the same as those in figure \ref{ipibranchsummary}. However, the different location of branch cut forces the starting valley $V_{1,0}$ to be shifted to its copy $V_{1,-1}$, a distance $2\pi i$ below in the $w$ integration plane. In other words, the principle sheet for the $-i\pi$ cut has been shifted and differs from that of the $+i\pi$ cut, as can be seen by comparing figures \ref{ipibranchsummary} and \ref{mipibranchsummary}.
At points e) the two saddles $w_{1,-1}$ and $w_{3,0}$ contribute (again corresponding to $\phi_{-k}^{\textrm{in}}$ and $\phi_{k}^{\textrm{in}}$, respectively), in contrast to the four contributions from the analogous $(z,\omega)$ point for the $+i\pi$ branch cut. This describes the wave interference we see in the right plot of figure \ref{fig:loggedHawking}. At the caustic d), two of these saddles $w_{1,-1}$ and $w_{2,-1}$ coalesce, so that there is one double saddle and one simple saddle contribution. Above the caustic at c) a Stokes's phenomenon is continuously occurring between $w_{1,-1}$ and $w_{2,-1}$, ($w_{1,-1}$ is subdominant to $w_{2,-1}$). This persists until the horizon at $z=0$ is the reached, at which point a double Stokes's phenomenon takes place at point b) (two dog leg turns on the steepest path from $w_{1,-1}$ to $w_{2,-1}$ to $w_{3,0}$), with $w_{3,0}$ dominant, $w_{2,-1}$ subdominant and $w_{1,-1}$ sub-subdominant. Finally, beyond the horizon at points a), the contour over $w_{1,-1}$ ($\phi_{\downarrow}^{\textrm{out}}$) passes onto the next lowest sheet, running into and out of $V_{3,-2}$ before encountering $w_{3,-1}$ ($\phi_{k}^{\textrm{out}}$) turns through a right angle before running back up to the principal sheet, passing over $w_{3,0}$ before running into $V_{3,0}$.  From this we observe that $w_{3,-1}$ is always undergoing a Stokes's phenomenon above the horizon for $z>0$, and is subdominant when compared to $w_{3,0}$.

We make the following remarks: First, the two choices of branches $+i\pi$ and $-i\pi$ display complementary contributions from sub-subdominant saddles located outside their respective principal Riemann sheets. Whenever the $+i\pi$ contour diagrams have contributing saddles outside the principal sheet the $-i\pi$ diagrams do not, and vice versa. 

Second, at points a), aside from the a single real wave, we have two additional waves: the evanescent wave from within the principal sheet and the sub-subdominant saddle from an adjacent one. This differs from the results of CPF \cite{coutant12}, where they only find there to be a single real saddle. This is because of their exclusion of the cubic term from the phase ($k_{c}\rightarrow\infty$), prior to applying the dominated convergence theorem (see equation 66 in \cite{coutant12}). The two approaches would agree if exponentially subdominant asymptotic contributions were to be neglected in the presence of more dominant ones. 

Third, along the horizon at $z=0$, a double Stokes's phenomenon takes place, as three saddlepoints are simultaneously joined by a single steepest path. This is also an indication of the potential for a higher order Stokes's phenomenon \cite{Howls04}, which would lead to additional interesting behaviour in the non-physical complex $(z,\omega)$ space. The horizon is indeed a part of a Stokes set (intersecting the already-identified real $(z,\omega)$ Stokes surface), no matter the choice of cut. 

\begin{figure}[t]
        \centering	
        \includegraphics[width=15cm]{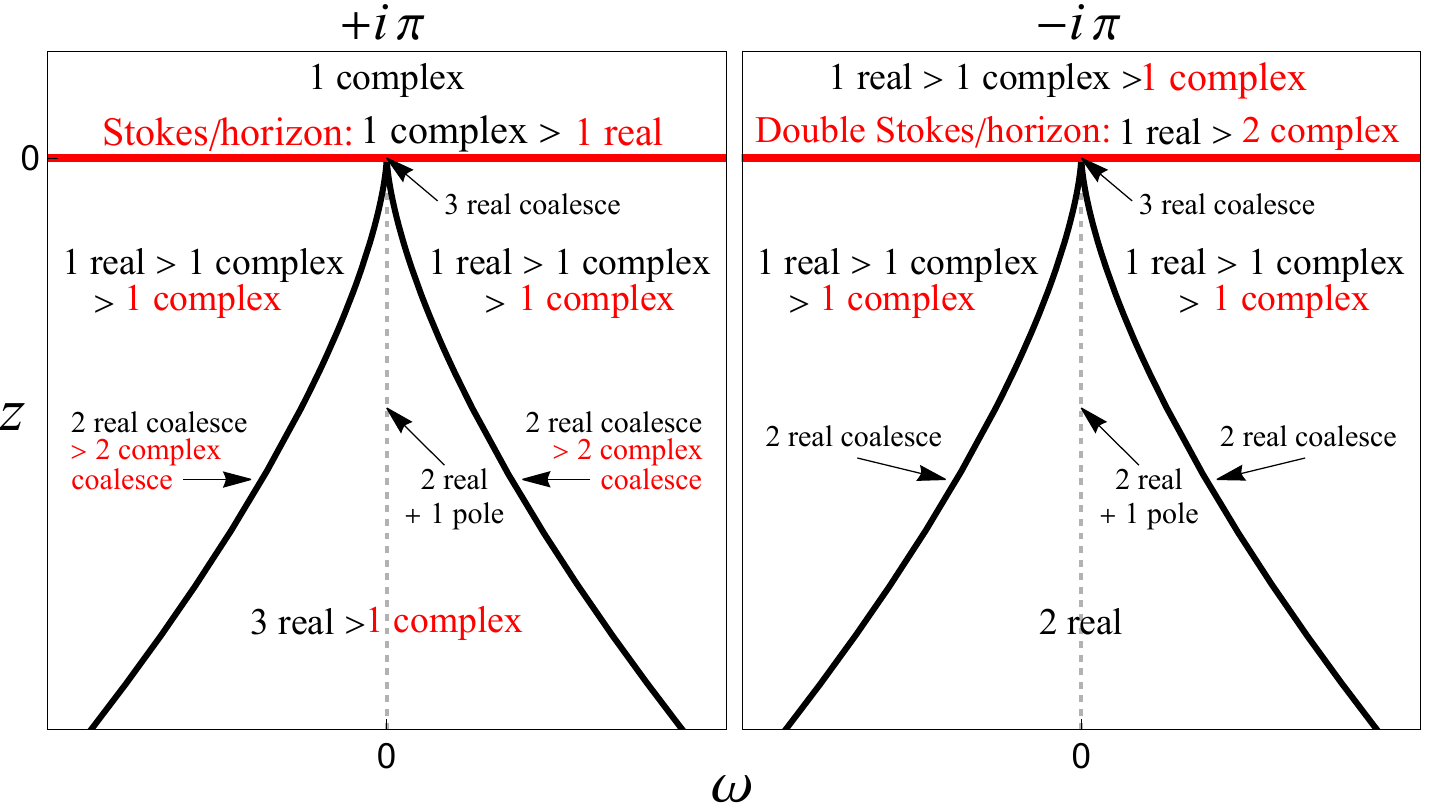} 
        \caption{The qualitative behavior of real and complex saddles $w_{j,n}$ of the log-Airy function Eq.\ (\ref{loggedHawking}) for both $+i\pi$ (\textbf{left}) and $-i\pi$ (\textbf{right}) choices of branch cut. Their relative dominance is denoted by the “$>$” signs. Black lines represent the caustic while red text or lines denote saddles/points in the ($z,\omega$) plane that are a part of the Stokes set. The light grey dashed line denotes where $\omega=0$ and there is a pole contribution, as previously discussed in the context of figure \ref{omega=0summaryfig}.}\label{loggedHawkingRegions2}
    \end{figure}

The overall qualitative behavior of the contributing saddlepoints (waves) is summarized in figure \ref{loggedHawkingRegions2} for both choices of cut, and describes the wave behavior observed in figures \ref{fig:loggedHawking}. The number of contributing real and complex saddlepoints together with their relative dominance throughout the $(z,\omega)$ plane is shown. Red lines and text denotes contributing saddles which are a part of the Stokes set, whether it be from within or outside of the principle Riemann sheet. The dominance of the saddles is denoted by the $``>"$ signs. These plots can be compared against figure \ref{pearceyPlot} for the Pearcey function. Clearly, the event horizon catastrophe has considerably more structure.\\

\subsection{Numerical Comparison and Validity of Asymptotics}

   \begin{figure}[htbp]
        \centering
        \captionsetup{width=1\linewidth}
        \includegraphics[width=14cm]{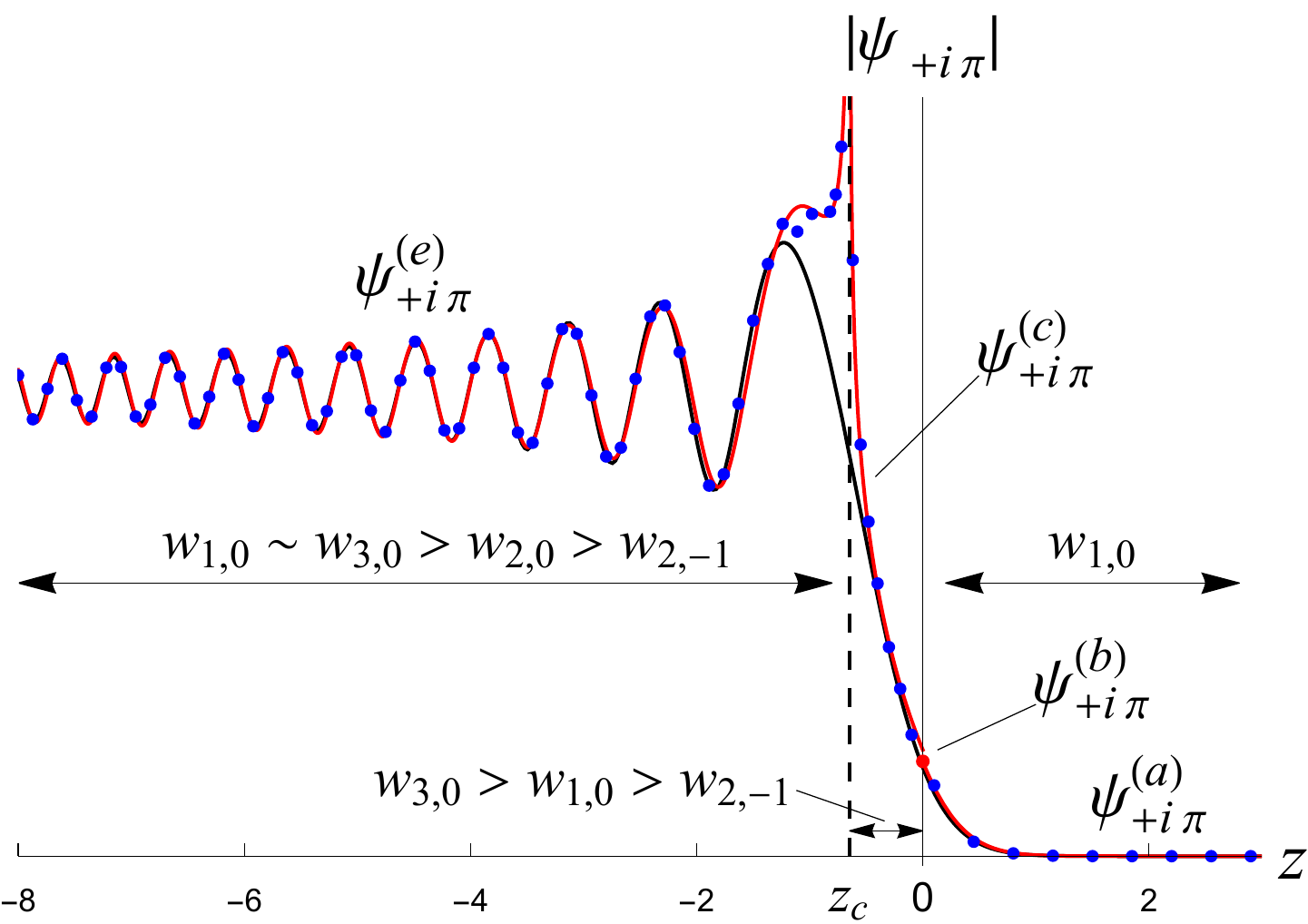} 			\caption{Numerically obtained exact plot (solid black line) of the log-Airy function Eq.\ (\ref{loggedHawkingscaled}) for the $+i\pi$ choice of branch cut with $\omega=1/5$ and $\lambda=5$. This is simply a $\omega>0$ slice of the lefthand plots in figure \ref{fig:loggedHawking}. The zeroth $r=0$ order asymptotic approximations a) through e) (solid red lines) and optimally truncated asymptotics (blue dots, see Appendix A) are also plotted for comparison. The red dot at $z=0$ corresponds to the asymptotic expression $\psi^{(b)}_{+i\pi}$ which is only valid at the horizon, and we do not label $\psi^{(d)}_{+i\pi}$ since point d) corresponds to the caustic where the asymptotic approximation diverges. As expected, the asymptotic expansions provide an (exponentially) good approximation to the exact result except near the caustic at $z_{c}=-3/(10^{2/3})$. The slight disagreement in the asymptotics at the horizon will vanish for large $\lambda\rightarrow\infty$. The saddles $w_{j,n}$ that the steepest paths encounter in each region are also indicated, together with their relative dominance (denoted by ``$>$"). The saddles $w_{1,0}$ and $w_{3,0}$ have equal real parts for $z$ below the caustic (denoted by ``$\sim$").} \label{comparisonfigure}  
    \end{figure}

    \begin{figure}[htbp]
        \centering
        \captionsetup{width=1\linewidth}
        \includegraphics[width=14cm]{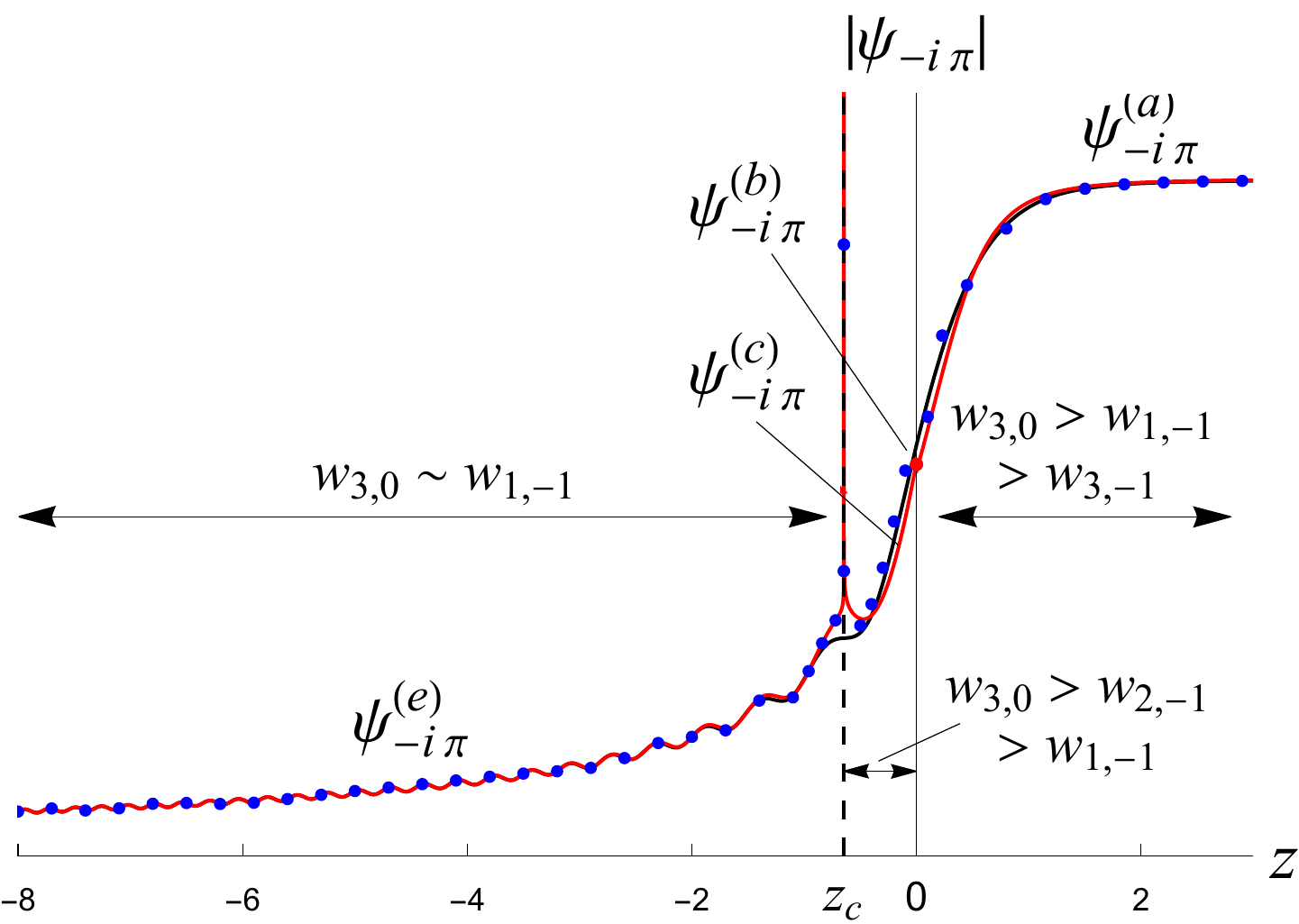} 	\caption{Numerically obtained exact plot (solid black line) of the log-Airy function Eq.\ (\ref{loggedHawkingscaled}) but now for the $-i\pi$ choice of branch cut. The notation and selected parameters are the same as those used in figure \ref{comparisonfigure}, and similarly this represents a constant $\omega>0$ slice of the righthand plots in figure \ref{fig:loggedHawking}. The zeroth $r=0$ order asymptotic approximations a) through e) (solid red lines) and optimally truncated asymptotic approximations (blue dots, see Appendix A) are again plotted for comparison. As expected, the asymptotic expansions provide an (exponentially) good approximation to the exact result except near the caustic at $z_{c}=-3/(10^{2/3})$ and exactly at the horizon $z=0$. The saddles $w_{1,-1}$ and $w_{3,0}$ have equal real parts for $z$ below the caustic (denoted by ``$\sim$").} \label{comparisonfigureMinPi}  
    \end{figure}

In figures \ref{comparisonfigure} and \ref{comparisonfigureMinPi} we demonstrate that our exponential approach  yields an accurate approximation  by plotting the asymptotic expressions against the exact wave functions $\psi_{\pm i \pi}$ obtained by numerically solving the integral in Eq.\ (\ref{loggedHawkingscaled})   for the two choices of branch cut. The plots are for $z$ lying in the range $-8<z<3$ and have the parameter values $\omega=1/5$ and $\lambda=5$. The solid black curves in the figures give the exact results whereas the red curves are composed of the asymptotic expressions given in Eqns.\ (\ref{asympexp})--(\ref{zeroTerm}) applied to the spatial points a) through e) [the latter of which we have explicitly studied for the $+i\pi$ cut and is given by Eq.\ (\ref{e-sum2})]. In the figures we have also included the information about the relevant contributing saddlepoints $w_{j,n}$ and their dominance for each of the spatial regions. Unlike CPF \cite{coutant12}, we have not made any small $\omega$ or $k_{c}\rightarrow\infty$ approximations. 

In making the red curves in figures \ref{comparisonfigure} and \ref{comparisonfigureMinPi} we have used only the zeroth order terms $r=0$ in the asymptotic expansions [see Eq.\ (\ref{asympexp})] and yet find an excellent match to the exact results (sufficiently far from the caustic). This is despite the fact that our `large' parameter $\lambda$ is only of order unity. The blow up close to the caustic could be fixed by using a uniform approximation \cite{DLMF,berry023_1972}.   
 It is also noteworthy that the asymptotic approximations for the different spatial regions match together so well at their respective borders. The only slight discrepancy is at the horizon between the asymptotic values of $\psi^{(a)}_{+i\pi}$, $\psi^{(b)}_{+i\pi}$ (the lone red dot in figure \ref{comparisonfigure}), and $\psi^{(c)}_{+i\pi}$. We find that when we use larger values of $\lambda$ this difference vanishes as expected so that $\psi^{(a)}_{+i\pi} \sim \psi^{(b)}_{+i\pi} \sim \psi^{(c)}_{+i\pi}$ at $z=0$ as $\lambda\rightarrow\infty$.

Eq.\ (\ref{asympexp}) is a diverging series, so there is an optimal series truncation in $r$ which can be made for each saddlepoint contribution, giving the approximation of lowest possible error. Taking terms higher than this value of $r$ will actually decrease the validity of the approximation and increase the error. Although we find that the lowest order $r=0$ truncation already gives a good visual match, we have also evaluated the higher terms and these are included as the blue dots in figures \ref{comparisonfigure} and \ref{comparisonfigureMinPi}. The details of the calculations of the higher order terms are given in Appendix A of this work.\\

\section{Concluding remarks}

In this paper we take a `catastrophe theory' approach to horizons motivated by the observation that a nonlinear dispersion relation causes the solutions of Hamilton's equations to undergo a broken pitchfork bifurcation near the horizon. Pitchfork bifurcations are specified by two control parameters which in our case are the lab frame frequency $\omega$ and the position coordinate $z$. 

According to catastrophe theory, the universal structurally stable relationship between these control parameters gives a cusp shape $z \propto - \omega^{2/3}$ which defines the location of a caustic $z_{c}$ where waves coalesce. However, whereas cusp caustics are usually dressed by the Pearcey function wave pattern, the event horizon bifurcation gives rise to a novel form of wave pattern described by an Airy function modified by a logarithmic term we call the log-Airy function. 

Some familiar properties remain such as self-similar scaling and we use this to identify a classical limit with a linear dispersion. Furthermore, like the Pearcey function there is a Stokes set that occurs outside the cusp, although in the log-Airy case it is flattened into a straight line in the ($\omega,z$) plane which coincides with the event horizon. 
Except for the special case of $\omega=0$, the caustic at $z_{c}$ [point d) in figure \ref{saddRegions}] and event horizon at $z=0$ [point b) in figure \ref{saddRegions}] do not sit at the same location: the caustic lies downstream behind the horizon  and the shape of the caustic implies that the spatial gap between them grows as $\omega^{2/3}$. This scaling has been pointed out before on the basis of the behaviour of the classical solutions \cite{coutant12}, and the connection between caustics and horizons was previously studied in a different way in the context of water waves \cite{Rousseaux2008,nardin2009}. However, the knowledge that it is a universal prediction of catastrophe theory and corresponds to the zone between a Stokes line and caustic adds to our understanding of the notion of a broadened horizon. On the other hand, this challenges us to generalize wave catastrophe theory to include logarithmic terms that ultimately arise from particle creation in quantum field theory \cite{leonhardt02,berry364_2004,berry401_2008}.

The log-Airy function has previously been analyzed by CPF in \cite{coutant12}. However, our treatment differs in some important respects. In CPF \cite{coutant12} the expansion in the region characterized by point e) (see figure \ref{saddRegions}) is based on two saddlepoints plus a loop contour around a branch cut. The effective removal of the logarithm by CPF from the phase significantly simplifies expressions for their two saddlepoint contributions, but restricts the validity of their results to $\vert z \vert \gg d$ and $\vert z \vert \gg d \, (\omega/\kappa)^{2/3}$ (units restored), see equations 34 and 57 in \cite{coutant12} and equations 11a and 11b in \cite{coutant14}. Their two saddlepoint contributions are equivalent to our Eqns.\ (\ref{T1}) and (\ref{T3}), which were obtained by performing a small $\omega$ expansion to our large $\lambda$ asymptotic result, Eq.\ (\ref{e-sum2}). In fact keeping the logarithm in the exponent is necessary to understand the role the latter plays as a Stokes line across which the number of contributing saddlepoints (and so waves) changes. The result of their loop contribution (corresponding to our Eq.\ (\ref{T2}) in the large $\lambda$ limit) is a complicated cut expansion involving the gamma function on its complex Stokes line. Their expression approximates $k_{c}\rightarrow\infty$, ignoring nonlinear effects, and obscures the underlying simplicity of the contributing subdominant copies of saddlepoints on adjacent Riemann sheets.

Although we only explicitly focused on an analytic description for points e) [Eq.\ (\ref{e-sum2})] for the $+i\pi$ choice of branch cut, our method was applied to the remaining points for both choices of cut and for a small constant value of $\omega>0$ in order to show the validity of our approach. This is shown in figures \ref{comparisonfigure} and \ref{comparisonfigureMinPi} for both the zeroth order analytic approximations [obtained via Eq.\ (\ref{asympexp})] and for the optimally truncated numerical asymptotics (described in Appendix A). Again, this approach could just as well be applied for larger $\omega$ and for $\omega<0$.

Our asymptotic contributions still diverge at the caustic as expected, but a uniform approximation \cite{DLMF,berry023_1972} could be employed locally to deal with this. Due to the divergence being attributable to a coalescence of two saddlepoints, this would take the form of an Airy function and its derivative. The caustic at which this occurs in the $(z,\omega)$ plane is one dimensional and hence even though there is a logarithm perturbing the polynomial in the exponent of the integral, the local behaviour across this cusp is still structurally stable and so falls within the realm of catastrophe theory.

Our large parameter $\lambda$ given in Eq.\ (\ref{eq:lambda_defn}) is a constant that does not depend on position, unlike $\Delta(z)$ defined in equation 34 of \cite{coutant12} which vanishes at the horizon. Combined with our transformation to exponential coordinates, this allows the near horizon behaviour to be examined with exponential accuracy, including the elucidation of new sub-subdominant contributions in regions below the caustic. Furthermore, the present catastrophe motivated approach could allow tight bounds to be put on corrections to the Hawking spectrum of emitted particles due to nonlinear dispersive effects, particularly if $\omega$ is not small. In particular, CPF \cite{coutant12} show how to combine the wave functions $\psi_{\pm i \pi}$ to obtain the Bogoliubov coefficients that directly give the Hawking production rate. This will be pursued in future work.

\section*{Acknowledgements}
The authors would like to thank two anonymous referees for their valuable feedback and suggestions. They also gratefully acknowledge the Isaac Newton Institute (INI) of Cambridge University for hosting the Applicable Resurgent Asymptotics Programme where part of this work was undertaken, and the Natural Sciences and Engineering Research Council of Canada (NSERC) for funding DO and LF. 

CH and DO would like to express their deep gratitude to Sir Michael Berry for his scientific and personal mentorship over many years. We believe we speak for generations of students and visitors in thanking both Michael and Monica Berry for providing a welcoming and nourishing environment in Bristol. We fondly remember our first encounters with wave singularities not just on paper but in real life in the form of expeditions to see the Severn bore. Chasing across the Gloucestershire countryside in Michael's car, sometimes at night, to an ideal viewing spot we recall the low roar of the wave as it approaches, which is especially intimidating after dark, accompanied by reversal of the river's flow direction, change in depth and sometimes wet feet. It provides observers with an impressive and tangible demonstration of Nature's power, not to mention the universality of the Airy function.

\appendix
\section{Numerical Evaluation of higher order Asymptotic Contributions}

The  evaluation of the higher order asymptotic contributions takes some care and we shall do it numerically.
Generically the terms in the asymptotic expansion Eq.\ (\ref{asympexp}) will decrease in magnitude, before diverging as $N_{j,n} \rightarrow \infty$. Therefore it is necessary to truncate each series at a finite value of $N_{j,n}$. This could also be done analytically via Eqns.\ (\ref{asympexp})--(\ref{termint}), but at the cost of some complexity.

In any practical numerical evaluation, the finite values of the $N_{j,n}$ in each saddlepoint expansion will depend on the relative sizes of the minimum values of $|T^{(j,n)}_r(z,\omega)|$.  For large $r$ these terms formally have the following expansion:
\begin{equation}
   T^{(j,n)}_r(z,\omega)\sim \sum_{l,p}\frac{K_{jl}}{2\pi{\rm i}}\sum_{s=0}^{+\infty}\frac{\Gamma(r-s)}{F_{jl}^{r-s}}T^{(l,p)}_s(z,\omega)\sim \frac{K_{jl}}{2\pi{\rm i}}\frac{\Gamma(r)}{\left(F_{j*l*}\right)^{r}T^{(l,p)}_0(z,\omega)}, \qquad r\rightarrow +\infty
\end{equation}
where the singulant $F_{jl}\equiv f_{j,n}-f_{l,p}$ is the difference in complex heights between the adjacent (sub) set of all saddlepoints which contribute to the expansion, and $K_{jl}$ is a Stokes constant \cite{Dingle}. The term $F_{j*l*}$ is the smallest in magnitude of all the singulants of the adjacent saddles \cite{BerryHowls}.

Let the set of saddles $w_{j,n}$ contributing at a point $(z,\omega)$ be $\mathcal W$, and let $r_{\rm min}$ be the index of the least term of each series $(j,n)$, defined by 
\begin{equation}
    |T^{(j,n)}_{r_{\rm min}(j,n)}(z,\omega)|=\min_{r}|T^{(j,n)}_r(z,\omega)|  \ .
\end{equation}
Furthermore, we define the largest in magnitude of each of these least terms as
\begin{equation}
    |T^{(j*,n*)}_{r*}(z,\omega)| = \max\{\min_{j,n}|T^{(j,n)}_r(z,\omega)|\}, \qquad r*=|\lambda F_{j*l*}|\qquad w_{j,n}\in {\mathcal W}. 
\end{equation}
Then for each $\psi^{(j,n)}_{r}(z,\omega)$ in Eqn.\ (\ref{asympexp}) only those terms which satisfy
\begin{equation}
    |T^{(j,n)}_r(z,\omega)| \ge |T^{(j*,n*)}_{r*}(z,\omega)|, \qquad  r<r_{\rm min}(j,n)
\end{equation}
should be included in the sum. This will lead to an approximation that is exponentially accurate to $\mathcal O(e^{-r*})$ \cite{BerryHowls,Dingle,Adri}.

\begin{figure}
    \centering
    \captionsetup{width=1\linewidth}
    \includegraphics[width=12cm]{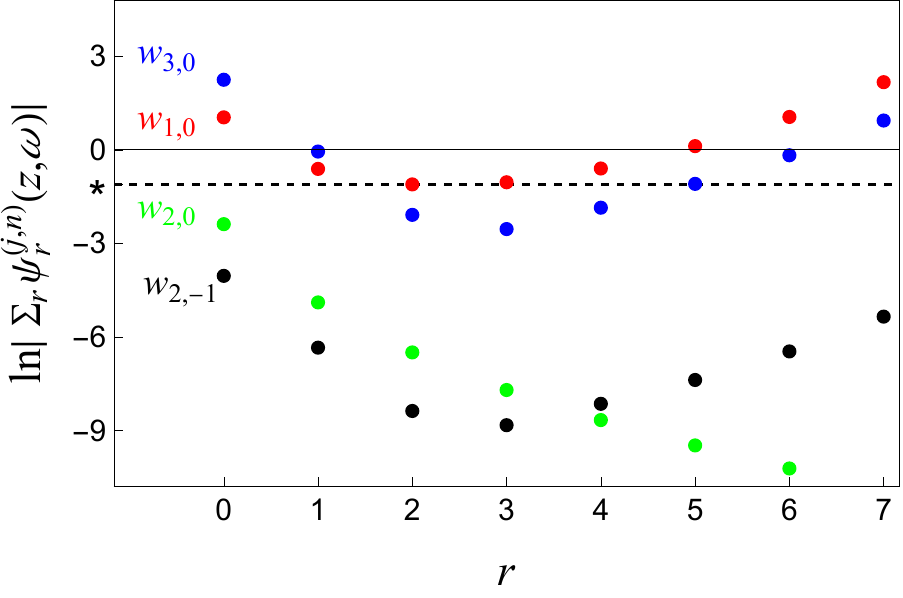}
    \caption{The relative magnitude of terms $\ln\left|\sum_{r}\psi_{r}^{(j,n)}(z,\omega)\right|$ in expansions about the saddle $w_{j,n}$ at $(z,\omega)=(-13/10,1/5)$, $\lambda=5$.  The maximum least term occurs at $r=2$ for the expansion about $w_{1,0}$ as indicated by the horizontal dashed line and denoted by the asterisk.}
    \label{termsize}
\end{figure}

Inclusion of terms in the divergent tail larger than the minimum only leads to increases in the inaccuracy of the approximation.  Inclusion of terms less than the size of the largest (in modulus) least term will not improve the accuracy, and likely lead to inaccuracy, since they ignore the (potentially resummable) contributions from the divergent tail after that largest term. 

To illustrate this point, figure \ref{termsize} shows the relative size of the terms for $(z,\omega)=(-13/10,1/5)$ [e) in figure \ref{saddRegions}], $\lambda=5$, in each of the four saddlepoint contributions $w_{1,0}$, $w_{2,0}$, $w_{2,-1}$, and $w_{3,0}$. The maximum of the least term in each set of terms is that of $w_{1,0}$ at $r=2$. Hence at this value of $(z,\omega)$, $T^{(j*,n*)}_{r*}(z,\omega)=T^{(1,0)}_{2}(z,\omega)$. The (exponentially small) error made in truncating at this term will be around the level of the horizontal dotted line. The terms in the expansions about $w_{2,0}$ and $w_{2,-1}$ are all smaller in magnitude than this error for all $r$ and so are numerically negligible. Therefore only the first three terms $r=0,1,2$ for $w_{1,0}$ and the first two terms $r=0,1$ for $w_{3,0}$ should be taken to contribute in the numeric asymptotic expansions. 

Hence, depending on the value of $(z,\omega)$ although topologically a saddle may contribute, if only the finite (yet exponentially accurate) sum to the maximum least term is taken, that saddle may not do so numerically.  
If a hyperasymptotic approximation \cite{BerryHowls} or a Borel-Pad{\'e} approach \cite{Ines} were to be undertaken in order to achieve a better than exponentially accurate approximation, then both categories of these neglected terms would need to be included.  However that is beyond the scope of this paper.

\section*{Bibliography}
\bibliographystyle{unsrt}

\end{document}